\documentclass[a4paper]{article}
\usepackage[utf8]{inputenc}
\usepackage[margin=1in]{geometry}     
\usepackage[parfill]{parskip}  
\usepackage{graphicx}
\usepackage{amssymb}
\usepackage{braket}
\usepackage{epstopdf}
\usepackage{mathtools}
\usepackage{subfig}
\usepackage[toc,page]{appendix}
\usepackage{cases}
\usepackage{tikz} 
\usetikzlibrary{arrows} 
\usetikzlibrary{shapes,arrows,positioning,automata,backgrounds,calc,er,patterns}
\tikzset{
    ->-/.style={decoration={
  markings,
  mark=at position .5 with {\arrow{>}}},postaction={decorate}},
    -<-/.style={decoration={
  markings,
  mark=at position .5 with {\arrow{<}}},postaction={decorate}},
    ->/.style={decoration={
  markings,
  mark=at position .4 with {\arrow{>}}},postaction={decorate}},
}
\usetikzlibrary{shapes,arrows,positioning,automata,backgrounds,calc,er,patterns}
\usetikzlibrary{calc,positioning,shadows.blur,decorations.pathreplacing}
\tikzset{%
        brace/.style = { decorate, decoration={brace, amplitude=5pt} },
       mbrace/.style = { decorate, decoration={brace, amplitude=5pt, mirror} },
        label/.style = { black, midway, scale=0.7, align=center },
     toplabel/.style = { label, above=.5em, anchor=south },
    leftlabel/.style = { label,rotate=-90,left=.5em,anchor=north },   
  bottomlabel/.style = { label, below=.5em, anchor=north },
        force/.style = { rotate=-90,scale=0.6 },
        round/.style = { rounded corners=2mm },
       legend/.style = { right,scale=0.6 },
        nosep/.style = { inner sep=0pt },
   generation/.style = { anchor=base }
}

\usepackage{tikz-feynman}
\usepackage{empheq}
\usepackage{slashed}
\usepackage{changepage}
\usepackage{jheppub}
\usepackage{fontawesome}
\newcommand{\ep}{\varepsilon}

\newcommand{\be}{\begin{equation}}
\newcommand{\ee}{\end{equation}}
\newcommand{\ba}{\begin{eqnarray}}
\newcommand{\ea}{\end{eqnarray}}
\newcommand{\Imag}{\mathop{\mathrm{Im}}}

\newcommand{\M}{ T}

\title{Unitarity and Unitarization}
\author{Alexandre Salas-Bernárdez}
\affiliation{Dept.  Análisis Matemático y Matemática Aplicada, Univ. Complutense de Madrid, Plaza de las Ciencias 3, 28040 Madrid, Spain.}
\emailAdd{alexsala@ucm.es}

\abstract{This article reviews unitarization methods essential for extending Effective Field Theories (EFTs) beyond their perturbative limits, particularly in hadronic and electroweak (EW) sectors. Perturbative EFTs, like Chiral Perturbation Theory (ChPT), often violate unitarity bounds at higher energies, a breakdown observed in phenomena such as $\pi\pi$ scattering resonances. To overcome this, non-perturbative techniques including the Inverse Amplitude Method (IAM), $K$-matrix formalism, and N/D approach are detailed. The IAM and the N/D methods resum perturbative series while preserving fundamental $S$-matrix principles: unitarity, analyticity, and causality, dynamically generating resonant behavior. The article emphasizes the unique role of dispersive frameworks, especially the Roy equations, which rigorously incorporate analyticity and crossing symmetry. It highlights their potential for future application in the electroweak sector, offering a powerful tool to constrain the Standard Model and interpret collider data.}

\begin{document}

\maketitle

\section{Introduction}\label{sec1}

Experimental data increasingly suggest that if new physics exists, it most likely resides at energy scales beyond the direct reach of current collider facilities, such that potential new resonances cannot be produced on-shell~\cite{ParticleDataGroup:2022pth}. Nevertheless, if these scales are not excessively high, their indirect effects may still manifest through the low-energy constants of an Effective Field Theory (EFT) built upon the known degrees of freedom of the Standard Model. In many cases, these constants multiply momentum-dependent operators, which can induce scattering amplitudes that grow rapidly with energy—eventually violating perturbative unitarity. This behavior is well known in hadronic physics and similarly arises in electroweak (EW) processes involving longitudinal gauge bosons.

A canonical example is Chiral Perturbation Theory (ChPT) (see \cite{Pich:1995bw}), which systematically describes the low-energy interactions among pseudo-Goldstone bosons (pGB) such as pions, kaons, and etas; as well as the scattering of EW scalar bosons(Higgs and pGB of the EW symmetry breaking) in what is known as the Higgs EFT (HEFT). Its expansion in powers of momenta is reliable only up to energies modestly above the lightest two-particle threshold. Beyond this range, perturbative predictions often breach unitarity bounds, thereby losing validity.

This breakdown is not merely formal; it is clearly observed in several physical processes. These include the resonant structure of $\pi\pi$ scattering phase shifts in the $J=0$ and $J=1$ partial waves~\cite{Ananthanarayan:2000ht,Garcia-Martin:2011iqs}, the appearance of broad scalar resonances such as the $f_0(500)$, and processes like $\eta \to 3\pi$~\cite{Gan:2020aco} or $\gamma\gamma \to \pi^0\pi^0$~\cite{Oller:2007sh}, where perturbative techniques fail to capture the full analytic and unitary structure of the $S$-matrix. Which we describe in sections \ref{sec:analytic}, \ref{sec:pwa} and \ref{sec:crossing}.

To extend the applicability of EFTs and recover unitarity beyond their naive range, several unitarization methods have been developed. Among them, well known is the \emph{Inverse Amplitude Method} (IAM), which enforces exact unitarity in elastic channels and is capable of predicting resonant behavior from low-energy input~\cite{Pelaez:2004xp,Oller:1997ng,Oller:1997ti}. The IAM is in close relation with the use of Padé approximants to partial wave amplitudes \cite{Willenbrock:1990dq}. The IAM, as well as other methods such as the N/D approach~\cite{Chew:1960iv,Oller:1998zr}, and various versions of the Bethe-Salpeter equation~\cite{Nieves:1999bx}, serve as effective tools to resum the perturbative series while preserving analyticity and causality. Another well known method is the K-matrix method which does explicitly violate the analyticity condition, producing spurious resonances in the physical sheet as seen in \cite{Truong:1991gv}. Each of these methods has its limitations as extensively studied in particular cases, see for example \cite{Truong:1991gv,Qin:2002hk,Masjuan:2008cp}. We will describe these unitarization methods in section~\ref{sec:pwaunit}. Other methods exist in the literature and will not be explained here, it is nonetheless interesting to mention the PKU method \cite{Zheng:2003rw,He:2002ut,Zhou:2006wm}, since it incorporates reliably unitarity and analiticity.

These approaches are not merely phenomenological; they are deeply rooted in the analytic structure of scattering amplitudes. In particular, dispersive unitarization techniques—discussed in section~\ref{sec:DispRel}—rely on dispersion relations to reconstruct amplitudes from first principles, using unitarity, analyticity, and crossing symmetry  (see sections \ref{sec:crossing} and \ref{subsec:crossingv}) as guiding constraints. This framework allows for the inclusion of model-independent information and provides a powerful way to access energy domains otherwise beyond reach. As a noteworthy example of these frameworks, we introduce in section \ref{sec:Roy} the Roy equations, which accurately include unitarity, analyticity, and crossing symmetry and are yet to be applied to the EW sector.

\section{Analytic Properties of Scattering Amplitudes}\label{sec:analytic}

Given a particle of four-momentum $p$, mass $m^2=p_\mu p^\mu$, three-momentum $\vec{p}$ and extra quantum numbers $\lambda$, its one-particle state $\lvert p \rangle = \lvert m, \vec{p}, \lambda \rangle$ is normalized as
\[
\langle p \vert p' \rangle = \langle m, \vec{p}, \lambda \vert m', \vec{p}\,', \lambda' \rangle = (2\pi)^3 2E_{\vec{p}} \, \delta^{(3)}(\vec{p} - \vec{p}\,') \, \delta_{mm'} \, \delta_{\lambda \lambda'}.
\]
The factor $2E_{\vec{p}}$ appears to make the normalization condition Lorentz invariant, and 
\begin{equation}
E_{\vec{p}} = \sqrt{\vec{p}^{\,2} + m^2}
\end{equation}
is the particle's energy. The multi-particle Fock space is defined using these single-particle states. An $n$-particle state is then defined as
\begin{equation}
\ket{m_1,\vec{p}_1, \lambda_1; ...;m_n,\vec{p}_n, \lambda_n}\equiv \ket{m_1,\vec{p}_1, \lambda_1}\otimes ... \otimes \ket{m_n,\vec{p}_n, \lambda_n}\;.
\end{equation}

For the case of $\pi\pi$ or $hh$ scattering, let us consider now the two-body scattering process of spinless particles $1 + 2 \rightarrow 3 + 4$ with masses $m_i$ for $i=1,...4$. The well-known Mandelstam variables are:
\begin{equation}
s = (p_1 + p_2)^2, \quad
t = (p_1 - p_3)^2, \quad
u = (p_1 - p_4)^2,
\end{equation}
with the relation:
\begin{equation}
s + t + u = \sum_{i=1}^{4} m_i^2.
\end{equation}
In the center of mass (CM) frame for this process we have
\[
p_1 = (E_1, \vec{p}\,), \quad p_2 = (E_2, -\vec{p}\,), \quad
p_3 = (E_3, \vec{p}\,'), \quad p_4 = (E_4, -\vec{p}\,'),
\]
and the Mandelstam variables are given by:
\begin{align}
s &= m_1^2 + m_2^2 + 2(E_1 E_2 + |\vec{p}\,|^2), \\
t &= m_1^2 + m_3^2 + 2(E_1 E_3 - |\vec{p}\,||\vec{p}\,'| \cos\theta), \\
u &= m_1^2 + m_4^2 + 2(E_1 E_4 + |\vec{p}\,||\vec{p}\,'| \cos\theta),
\end{align}
where $\theta$ is the angle between the three-momenta of particles 1 and 3 in the CM frame.

We can also express
\begin{align}
E_1 &= \frac{1}{2\sqrt{s}}(s + m_1^2 - m_2^2), \quad
E_2 = \frac{1}{2\sqrt{s}}(s + m_2^2 - m_1^2),\nonumber \\
E_3 &= \frac{1}{2\sqrt{s}}(s + m_3^2 - m_4^2), \quad
E_4 = \frac{1}{2\sqrt{s}}(s + m_4^2 - m_3^2),\nonumber \\
|\vec{p}\,|\, &= \frac{1}{2\sqrt{s}} \sqrt{(s - (m_1 + m_2)^2)(s - (m_1 - m_2)^2)}, \nonumber\\
|\vec{p}\,'|\, &= \frac{1}{2\sqrt{s}} \sqrt{(s - (m_3 + m_4)^2)(s - (m_3 - m_4)^2)}.\label{eq:momenta}
\end{align}

We can see from the expressions for the Mandelstam variables above that a physical process requires:
\[
s \geq (m_1 + m_2)^2.
\]
in this case we say that $s$ is above the production threshold.
For equal masses $m_i = m$, $i = 1, \ldots, 4$, such that $|\vec{p}\,| = |\vec{p}\,'|$, we get:
\begin{align}
s = 4(m^2 + |\vec{p}\,|^2),\;\;\;t = -2|\vec{p}\,|^2 (1 - \cos\theta),\;\;\;u = -2|\vec{p}\,|^2 (1 + \cos\theta).
\end{align}
In this case, the conditions for a physical process are
\begin{equation}
s \geq 4m^2, \quad 4m^2-s\leq t \leq 0, \quad  4m^2-s \leq u \leq 0. \label{eq:sregion}
\end{equation}
These conditions above will be regarded as the \textit{physical region} in what follows.

\paragraph{The $S$-matrix}
 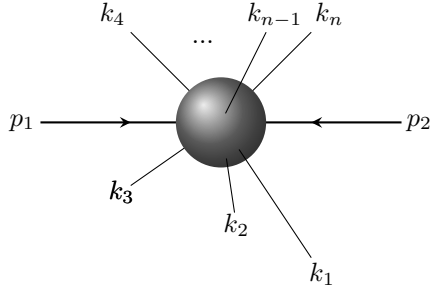
\begin{figure}[ht!]
     \centering
 \begin{tikzpicture}[>=stealth,scale=1.2]
  \draw[->-,thick] (-2,0)--(0,0);
    \draw[->-,thick] (2,0)--(0,0);
    
    \draw (0,0)--(1,1);
    \draw (0,0)--(-1,1);
    \draw (0,0)--(-1,-.7);
    \shade[ball color=gray]  (0,0) circle (.5cm);
    \draw (0.05,.1)--(10*0.05,10*.1);
    \draw (2*0.03,2*-.2)--(5*0.03,5*-.2);
    \draw (0.2,-.3)--(5*0.2,5*-.3);
    \draw (5.6*0.2,5.6*-.3) node {$k_1$};
    \draw (5.7*0.03,5.7*-.2) node {$k_2$};
    \draw (11.9*0.05,11.9*.1) node {$k_{n-1}$};
    \draw (1.1*-1,1.1*-.7) node {$k_{3}$};
    \draw (1.1*-1,1.1*-.7) node {$k_{3}$};
    \draw (1.2*-1,1.2*1) node {$k_{4}$};
    \draw (1.2*-1+1,1.2*1-.3) node {$...$};
    \draw (2.2,0) node {$p_2$};
        \draw (-2.2,0) node {$p_1$};
    \draw (1.2*1,1.2*1) node {$k_{n}$};
\end{tikzpicture}
     \caption[Representation of a $2\to2$ scattering process]{Representation of a scattering process where two incoming particles with momentum $p_1$ and $p_2$ collide producing $n$ outgoing particles travelling freely with momentum $k_i$ each.}
     \label{fig:collider}
 \end{figure}
The $S$-matrix is an operator that relates the Fock space of incoming particle states to the outgoing one (see \cite{Mizera:2023tfe} for a modern introduction): consider the probability that an initial state evolves under the time-evolution operator \( e^{-i\hat{H}t} \) into a given final state. Both the initial and final state particles are assumed to be accurately described by wave packets that are sharply peaked in momentum---that is, they possess well-defined momenta. The incoming state \( \ket{\vec{p}_1,\vec{p}_2 \, \text{in}} \) and outgoing state \( \ket{\vec{k}_1, \ldots, \vec{k}_n \, \text{out}} \) are defined such that the incoming particles are created at \( T \to -\infty \) and the outgoing particles are detected at \( T \to +\infty \). Under these assumptions, and ignoring the extra quantum numbers $\lambda_i$ and masses $m_i$ indices for clarity, their overlap is given by
\begin{align}
&\braket{\vec{k}_1, \ldots, \vec{k}_n \, \text{out} | \vec{p}_1, \vec{p}_2 \, \text{in}} =\lim_{t \to \infty} \braket{\vec{k}_1, \ldots, \vec{k}_n, +t | \vec{p}_1, \vec{p}_2, -t} \\
 &= \lim_{t \to \infty} \braket{\vec{k}_1, \ldots, \vec{k}_n | e^{-2i\hat{H}t} | \vec{p}_1, \vec{p}_2} \;,
\end{align}
where in the last expression all states are taken at a common reference time, \( t = 0 \). This limiting procedure defines the \emph{$S$-matrix}, whose matrix elements are
\begin{equation} \label{eq:Smatrixelements}
\braket{\vec{k}_1, \ldots, \vec{k}_n \, \text{out} | \vec{p}_1, \vec{p}_2 \, \text{in}} \equiv \braket{\vec{k}_1, \ldots, \vec{k}_n | \mathcal{S} | \vec{p}_1, \vec{p}_2} \;.
\end{equation}
This operator, $\mathcal{S}$, helps in computing the probability, $P_{fi}$, of an incoming multi-particle state $\ket{i {\text{ in}}}$ evolving into the outgoing multi-particle state $\ket{f{\text{out}}}$ by relating the two states to a common reference time
\[
P_{fi} = |{}_\text{out}\braket{f|i}_\text{in}|^2  = |{}\bra{f} \mathcal{S}\ket{i}|^2 = \langle i | {\mathcal{S}}^\dagger | f \rangle \langle f | {\mathcal{S}} | i \rangle.
\]
Where $\ket{i}$ and $\ket{f}$ are states in the common reference time. This expression gives the probability that the system is found in the final state $\lvert f \rangle$, given that it was initially in the state $\lvert i \rangle$, with both $\lvert i \rangle$ and $\lvert f \rangle$ belonging to the same Fock space.

Starting from an initial state $\ket{i}$, the probability of ending in all possible states $\ket{f}$ is, by definition of a probability, equal to one:
\[
\sum_f \lvert \langle f \vert \mathcal{S} \vert i \rangle \rvert^2 = \sum_f \langle i \vert \mathcal{S}^\dagger \vert f \rangle \langle f \vert \mathcal{S} \vert i \rangle = \langle i \vert \mathcal{S}^\dagger \mathcal{S} \vert i \rangle=1,
\]
where we have assumed that $\lvert f \rangle$ form a complete set of orthonormal states, so $\sum_f \lvert f \rangle \langle f \rvert = 1$ (this sum includes a phase space integration that we omit). Similarly, we could have proven that $\sum_f \lvert f \rangle \langle f \rvert = 1$.

Let us assume now that the initial state can be decomposed in terms of two orthonormal states (this can be generalized to any number of states, even the continuum)
\[
\lvert i \rangle = \frac{1}{\sqrt{| \alpha |^2 + | \beta |^2}} ( \alpha \lvert a \rangle + \beta \lvert b \rangle ).
\]
Using that $\braket{i|i}=1$ we obtain
\[
 \frac{1}{| \alpha |^2 + | \beta |^2} \left(\text{Re}(\alpha^* \beta \langle a \vert \mathcal{S}^\dagger \mathcal{S} \vert b \rangle) \right)=0.
\]
$\alpha$ and $\beta$ are arbitrary complex numbers and $\lvert a \rangle$, $\lvert b \rangle$ are arbitrary states. This means that off diagonal terms of the $\mathcal{S}\mathcal{S}^\dagger$ operator vanish, $\braket{a|\mathcal{S}^\dagger \mathcal{S}|b}=0$ for all orthonormal states $\ket{a}\neq \ket{b}$.

Consequently, the $S$ matrix operator is unitary:
\[
\mathcal{S} \mathcal{S}^\dagger = \mathcal{S}^\dagger \mathcal{S} = \mathbb{I}.
\]

 The interacting part of the $S$-matrix is referred to as the \emph{transition operator} \( \mathcal{T} \), and after factoring out total momentum conservation, the \emph{scattering amplitude} \(T \) is defined through
\begin{align}
\langle &\vec{k}_1 \mu_1; \ldots; \vec{k}_n \mu_n | (\mathcal{S} - \mathbb{I}) | \vec{p}_1 \lambda_1; \vec{p}_2 \lambda_2 \rangle \equiv \langle \vec{k}_1 \mu_1; \ldots; \vec{k}_n \mu_n | i\mathcal{T} | \vec{p}_1 \lambda_1; \vec{p}_2 \lambda_2 \rangle \nonumber \\
&\equiv i(2\pi)^4 \delta^4\left( \sum_{i=1}^n k_i - p_1 - p_2 \right) T(p_1\lambda_1, p_2\lambda_2 \to k_1\mu_1, \ldots, k_n\mu_n) \;.
\end{align}

This defines $\mathcal{T}$ as
\[
\langle f \vert \mathcal{S} \vert i \rangle = \delta_{if} + i \langle f \vert T \vert i \rangle \quad \Rightarrow \quad \mathcal{S} = \mathbb{I} + i\mathcal{T}.
\]

The unitarity of the $S$-matrix implies that:
\[
\delta_{ij} = \langle j \vert \mathcal{S} \mathcal{S}^\dagger \vert i \rangle = \sum_f \langle j \vert \mathcal{S} \vert f \rangle \langle f \vert \mathcal{S}^\dagger \vert i \rangle,
\]
where $\lvert i \rangle$, $\lvert j \rangle$ and $\lvert f \rangle$ are arbitrary orthonormal states. Using the definition of the transition matrix above, we obtain:
\[
\langle j \vert \mathcal{T} \vert i \rangle - \langle j \vert \mathcal{T}^\dagger \vert i \rangle = i  \sum_f \braket{j|\mathcal{T}|f}\braket{f|\mathcal{T}^\dagger|i}.
\]
Which is usually referred as the generalized \textit{optical theorem}. This theorem is closely related to the Largest Time Equation in coordinate-space \cite{Veltman:1963th}, which is a result of causality, and the Cutkowsky rules \cite{Cutkosky:1960sp} in momentum space.

For the particular case $j = i$ (which is the case for  $2\to2$ particle scattering), and assuming time reversal symmetry of the interactions, so that $ T_{fi} =  T_{if}$, we have:
\begin{equation}
2 \, \text{Im} \langle i \vert \mathcal{T} \vert i \rangle = (2\pi)^4 \sum_f \delta^{(4)}(p_f - p_i) {|T}_{fi}|^2,\label{eq:unitarity}
\end{equation}
where $ T_{fi}$ is the scattering amplitude for the process $\ket{i}\to \ket{f}$.

\section{Partial-wave expansion}\label{sec:pwa}
Conservation of wavefunction probability in  $2\to2$ scattering processes is expressed through a nonlinear integral relation satisfied by the scattering amplitude \( T_I(s,t,u) \), where \( s \), \( t \), and \( u \) are the Mandelstam variables, and \( I \) denotes the isospin (either strong isospin in hadron physics or electroweak isospin in Higgs Effective Theory). We will drop the isospin index from now on and introduce it back when useful.

Upon decomposing \( \M \) into partial waves of definite angular momentum \( J \), the unitarity condition takes a much simpler form, as seen in eq.~(\ref{OpticalTh}). The partial wave expansion reads:
\begin{equation}
    \M(s,t,u) = 16\eta\pi \sum_{J=0}^\infty (2J+1) \, t_{J}(s)\, P_J(\cos\theta) \,,\label{eq:16}
\end{equation}
which converges for physical values of \( s \) and for the scattering angle \( x \equiv \cos \theta_s \in [-1,1] \). Furthermore, convergence holds over the Lehmann ellipse for unphysical values of \( \cos\theta \), where the asymptotic behavior of the Legendre polynomials \( P_J \) remains under control~\cite{Lehmann:1972kv}. The symmetry factor is \( \eta = 2 \) if the scattering particles in the final and initial states are indistinguishable, \( \eta = \sqrt{2} \) if one two-particle state (final or initial) has distinguishable particles, and \( \eta = 1 \) otherwise. The inverse formula defining the partial waves is given by:
\begin{equation}\label{eq:pwa}
    t_{J}(s) = \frac{1}{32\pi\eta} \int_{-1}^{+1} \! dx \, P_J(x) \, \M(s,t(x),u(x)) \,.
\end{equation}

\par

When the final and initial states are identical, the partial waves inherit the analytic structure of the amplitude, which—due to crossing symmetry—also develops a left-hand cut (LC) along the negative real axis (see section \ref{sec:crossing} below). The LC originates from \( t \)- and \( u \)-channel exchanges of physical particles and is tied to unavoidable singularities of the amplitude at the endpoints \( x = \pm 1 \).

Inelastic channels involving multiparticle states such as \( 2n \pi \), \( K\bar{K} \), \( \eta\eta \), etc., in hadron physics—or \( 2n \, w_i \), \( hh \), etc., in the electroweak sector—introduce additional cuts starting at thresholds like \( (2n)^2 m_\pi^2 \), \( 4m_K^2 \), \( 4m_\eta^2 \), and so forth, extending to \( +\infty \).

\paragraph{Two-particle intermediate states}
 Consider the $s$-channel center-of-mass frame, where the initial and final states are
\[
|i\rangle = |p_1, p_2\rangle, 
\qquad 
|f\rangle = |p_3, p_4\rangle,
\]
with $\vec{p}_2 = -\vec{p}_1$ and $\vec{p}_4 = -\vec{p}_3$. Thus, both states have total momentum equal to zero and total energy $\sqrt{s}$.  

Restricting to two-particle intermediate states of the form $|n\rangle = |k_n, k_n'\rangle$, the unitarity condition for the transition amplitude in equation (\ref{eq:unitarity}), takes the form
\begin{align}
2\,&\text{Im } T(s,t,u) \nonumber=\\
&=(2\pi)^4 \sum_n 
\int \frac{d^3k_n}{2E_{k_n}(2\pi)^3} 
     \frac{d^3k_n'}{2E_{k_n'}(2\pi)^3} \,
\delta\!\big(E_{k_n}+E_{k_n'}-\sqrt{s}\big)\,
\delta^{(3)}(\vec{k}_n+\vec{k}_n') \,
T_{fn}T_{ni}^* .
\end{align}
Here the index $n$ labels only the type of intermediate state, since the integration over momenta is carried out explicitly.  

Performing the integrals using the delta functions, one obtains the solid angle integral
\begin{align}
\text{Im } T(s,t,u) 
= \,\frac{2|\vec{k}_f|}{64\pi^2\sqrt{s}} 
\sum_n \int d\Omega_{\vec{k}_n}\, 
T_{fn}T_{ni}^* ,\label{eq:unit1}
\end{align}
where the intermediate momentum $k_n$ satisfies eqns.~(\ref{eq:momenta}).  

The left-hand side of eq. (\ref{eq:unit1}) is, in terms of partial waves from eq.~(\ref{eq:pwa}),
\begin{align}
\sum_J (2J+1)\, P_J(\hat{p}\cdot\hat{p}') \,
\text{Im } t^{fi}_J(s),
\label{eq:LHS}
\end{align}
with $\hat{v}$ denoting the unit vector in the direction of $\vec{v}$. 

Inserting the partial-wave expansion of the amplitude, the right-hand side of eq.~(\ref{eq:unit1}) becomes
\begin{align}
4 \sum_n \frac{2|\vec{k}_n|}{\sqrt{s}} \sum_{J,J'}
(2J+1)(2J'+1)\, t^{fn}_J(s)\, t^{ni}_{J'}(s)^* 
\int d\Omega_{\vec{k}_n}\, 
P_J(\cos\theta_{fn})\,
P_{J'}(\cos\theta_{in}) ,
\end{align}
where the scattering angles are defined by
\[
\cos\theta_{in} = \hat{p}\cdot\hat{k}_n,
\qquad 
\cos\theta_{fn} = \hat{k}_n\cdot\hat{p}'.
\]

Using the spherical-harmonic expansion of the Legendre polynomials,
\begin{align}
P_J(\hat{p}\cdot\hat{k})
= \frac{4\pi}{2J+1} 
\sum_{M=-J}^J 
Y_{JM}^*(\hat{p})\,Y_{JM}(\hat{k}),
\end{align}
together with the orthonormality relation
\begin{align}
\int d\Omega_{\vec{k}}\,
Y_{JM}^*(\hat{k})\,Y_{J'M'}(\hat{k}) 
= \delta_{JJ'}\delta_{MM'} ,
\end{align}
one finds
\begin{align}
{16\pi}
\sum_J (2J+1)\, P_J(\hat{p}\cdot\hat{p}') 
\sum_n \frac{|\vec{k}_n|}{\sqrt{s}} t^{fn}_J(s)\, t^{ni}_J(s)^* .
\end{align}

Therefore, the unitarity condition for partial-wave amplitudes reads
\begin{align}
\text{Im } t^{fi}_J(s) 
= \sum_n \sigma_n(s)\, 
t^{fn}_J(s)\,t^{ni}_J(s)^* ,\label{eq:pwunit}
\end{align}
where
\[
\sigma_n(s) = \frac{2 |\vec{k}_n|}{\sqrt{s}}
\]
is the two-body phase space factor.  

\paragraph{Elastic scattering}
In the case of purely elastic scattering, i.e. below the first inelastic threshold, eq.~(\ref{eq:pwunit}) reduces to
\begin{align}
\text{Im } t_J(s) = \sigma(s)\, |t_J(s)|^2 ,\label{OpticalTh}
\end{align}
where \( \sigma(s) = \sqrt{1 - 4m^2/s} \) is the usual phase-space factor, which becomes unity for massless particles.

Hence, unitarity of the \( S \)-matrix implies that the partial waves satisfy a simple but nonlinear relation for physical values \( \text{Re}(s) > 0 \), analogous to the optical theorem.
This condition implies the bound:
\begin{equation}
    |t_{J}(s)| \leq \frac{1}{\sigma(s)} \,, \label{unitarity}
\end{equation}
for physical \( s \) above threshold.

\par
A central observation behind the Inverse Amplitude Method (see section \ref{subsec:IAM} below) is that, for purely elastic processes, eq.~(\ref{OpticalTh}) determines the imaginary part of the inverse of the partial wave:
\begin{equation}
    \text{Im} \left( \frac{1}{t_{J}(s)} \right) = -\sigma(s) \quad \text{for} \quad s > s_{\text{th}} \,. \label{unitarityinverse}
\end{equation}
This is an exact, non-perturbative result as long as inelastic channels (such as \( \pi\pi \to \pi\pi\pi\pi \) in hadron physics or \( W_L W_L \to W_L W_L W_L W_L \) in HEFT) can be neglected. 

\par
If inelasticity arises from an additional two-body channel—such as \( K\bar{K} \) in hadron physics or \( hh \) in the electroweak sector—the IAM must be extended to a matrix formulation \cite{GomezNicola:2001as}. This will be discussed in section \ref{subsec:coupledIAM}.

\par
Given that the imaginary part of the inverse amplitude is determined by kinematics alone, a natural question is how to recover its real part, which carries dynamical information. This question motivates the usage of Dispersion Relations, which will be covered in section \ref{sec:DispRel}.

\subsection{Three-body partial waves}

Three-body scattering amplitudes are essential for describing resonances that decay into three-hadron final states. Modern approaches extend the well-established two-body partial-wave expansion to $3 \to 3$ processes, relying on the foundational $S$-matrix principles of analyticity and unitarity~\cite{Mai:2017vot,Mikhasenko:2019vhk}. The general strategy involves decomposing the full amplitude $T_{3 \to 3}(s,t,u)$ into a sum of contributions from two-body subchannel (isobar) dynamics interacting with a spectator. For spinless particles, the amplitude is typically expressed as a cyclic sum over isobar--spectator terms,
\[
T = \sum_{i=1}^3 T^{(i)}(s; \sigma_i, \sigma'_i),
\]
where $\sigma_i$ denotes the invariant mass of the pair when particle $i$ is the spectator. Each component $T^{(i)}$ satisfies an integral equation that incorporates the two-body amplitudes and one-particle exchange (OPE) kernels. This construction guarantees exact three-body unitarity above the breakup threshold~\cite{Mai:2017vot}. By assuming that the exchanged particle propagates on shell, the full three-body Bethe--Salpeter or Faddeev equation can be reduced to a three-dimensional form. The resulting system enforces both three-body and subchannel unitarity for arbitrary energies, even in the presence of resonant two-body subsystems~\cite{Mai:2017vot,Mikhasenko:2019vhk}.

The equations can be cast as a coupled set of integral equations:
\[
T^{(i)} = V^{(i)} + \sum_{j \neq i} V^{(i)} G_0\, T^{(j)},
\]
where $V^{(i)}$ includes both the long-range OPE interaction and a short-range contact kernel, while $G_0$ is a product of three free-particle propagators with one on shell.

In this framework, the amplitude is decomposed in terms of two-body isobars interacting with a spectator, which naturally incorporates subchannel unitarity~\cite{Mai:2017vot}. The interaction kernel is often separated into a long-range part—arising from ladder-type OPE contributions—and a short-range component $R$, which encodes genuine three-body forces and irreducible contributions~\cite{Mikhasenko:2019vhk}. A factorization ansatz is frequently employed for the short-range term:
\[
R(\sigma,\sigma') = f(\sigma)\, g(\sigma'),
\]
which leads to a simple algebraic unitarity condition of the form
\[
R - R^\dagger = i R^\dagger \rho_3 R,
\]
where $\rho_3$ is the three-body phase-space factor. Notably, the ladder resummation of the OPE kernel is analytically equivalent to solving the Khuri--Treiman equations in a dispersive framework~\cite{Mikhasenko:2019vhk}, thereby establishing a direct correspondence between diagrammatic and dispersive $S$-matrix approaches.

\paragraph{Algebraic unitarity and integral equations}

Within the factorized representation, the connected amplitude $T_c$ can be written as the sum of a long-range part $L$ and a short-range term $R$,
\[
T_c = L + R,
\]
which satisfy coupled nonlinear equations:
\[
L = L\,\tau\,L, \qquad R - R^\dagger = R^\dagger (\tau - \tau^\dagger) R,
\]
where $\tau$ represents an effective phase-space integral operator~\cite{Mikhasenko:2019vhk}. 

Beyond the factorized scenario, fully relativistic Faddeev-type integral equations can be formulated and projected into definite angular momentum components. Jackura and Brice\~{n}o~\cite{Jackura:2023qtp} develop a relativistic representation where the amplitude is expanded in terms of the total angular momentum $J$ using Legendre polynomial projections:
\[
\mathcal{L}_J(s;\sigma,\sigma') = \frac{1}{2} \int_{-1}^{1} d\cos\theta\, P_J(\cos\theta)\, \mathcal{L}(s;\sigma,\sigma',\cos\theta),
\]
with $\mathcal{L}$ the OPE kernel defined in the isobar rest frame. These projections are crucial for ensuring that singularities are under control and for enabling robust numerical implementations. This formalism lays the foundation for analyzing hadronic three-body resonances and is directly applicable to both phenomenological analyses and lattice QCD studies.

\section{Crossing and analyticity}\label{sec:crossing}
Let us consider the  $2\to2$ particle scattering, $1+2\rightarrow 3+4$, and its scattering matrix, $ T_{1+2 \rightarrow 3+4}(s, t, u)$ in terms of the Mandelstam variables. The \textit{Mandelstam hypothesis} \cite{Mandelstam:1958xc} assumes that the scattering amplitude, $ T_s(s, t, u)$ in the so called $s$-region or physical region (defined by the inequalities in (\ref{eq:sregion})) can be analytically continued to the so called $t$-region
\begin{equation}
t \geq 4m^2, \quad 4m^2-t \leq s \leq 0, \quad 4m^2-t \leq u \leq 0, \label{eq:tregion}
\end{equation}
and $u$-region
\begin{equation}
u \geq 4m^2, \quad  4m^2-u \leq t \leq 0, \quad 4m^2-u \leq s \leq 0, \label{eq:uregion}
\end{equation}
in a unique way.

We can see the $t$-channel process as being:
\begin{equation}
1 + \bar{3} \rightarrow \bar{2} + 4, \label{eq:tcha}
\end{equation}
where the bars indicate that we deal with antiparticles. So that the $t$-region of the original process is the $s$-region for this process in eq. (\ref{eq:tcha}),
\begin{equation}
 T_{t}(t, s, u)\equiv T_{1 + \bar{3} \rightarrow \bar{2} + 4}(t, s, u) =  T_{1+2 \rightarrow 3+4}(s, t, u). 
\end{equation}

Similarly, for the $u$-channel process:
\begin{equation}
1 + \bar{4} \rightarrow \bar{2} + 3, 
\end{equation}
we have:
\[
 T_{u}(u, t, s)\equiv T_{1 + \bar{4} \rightarrow 3 + \bar{2}}(u, t, s) =  T_{1+2 \rightarrow 3+4}(s, t, u).
\]

\begin{figure}[ht!]
 \centering
\includegraphics[width=0.6\columnwidth]{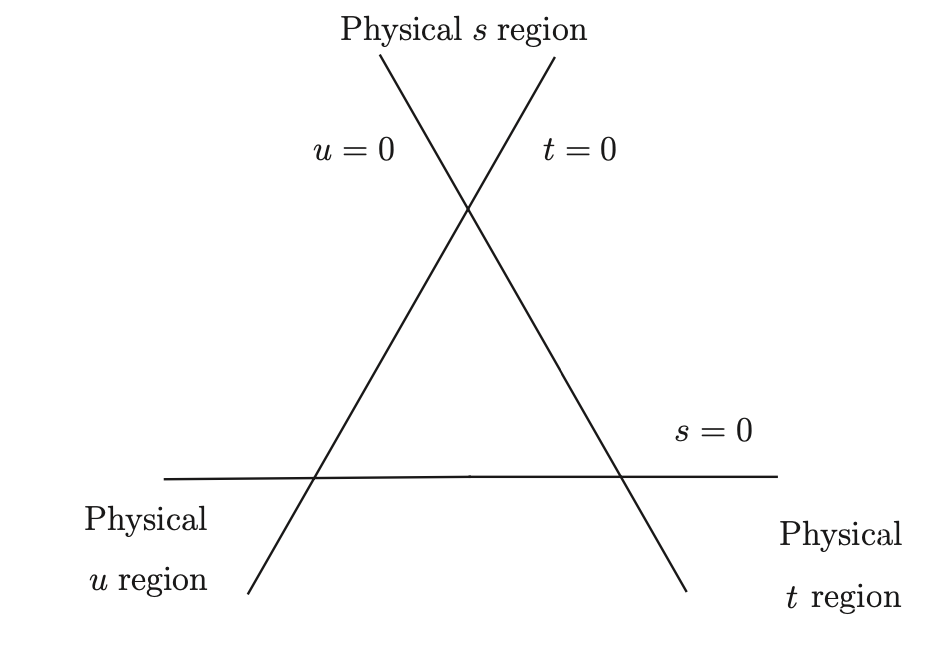}
   \caption[Mandelstam plane for the scattering of two identical particles]{Mandelstam plane for two identical particles.}\label{regionm}
 \end{figure}

Thus, Mandelstam's hypothesis assumes that there is a unique analytic amplitude $ T(s,t,u)$ such that
   \begin{align}
 T(s,t,u)=    \left\{
   \begin{array}{l}
 T_s(s,t,u) \,\, \text{for}\,\,s\geq4m^2,\,t\leq0,\,u\leq0\ ,\nonumber\\
 T_{t\,}(t,s,u) \,\, \text{for}\,\,t\geq4m^2,\,s\leq0,\,u\leq0\ ,\nonumber\\
  T_u(u,t,s) \, \text{for}\,\,u\geq4m^2,\,t\leq0,\,s\leq0\ ,
              \end{array}
    \right.
  \end{align}
  with these three physical regions of different channels 
  depicted Fig.~\ref{regionm} .
  
The Mandelstam hypothesis does not suffice to characterize fully the scattering amplitude and typically there are some extra assumptions to take. Usually, bound states appear as poles of the scattering amplitude in the real axis of the $s$ variable complex plane. Particle thresholds (\textit{i.e.} the energy of the process is enough to generate on-shell intermediate states) give rise to cuts due to the optical theorem in eq. (\ref{eq:unitarity}), since the amplitudes $ T_{fi}$ become non-zero.  In that equation, \( p_f^2 = s \) denotes the invariant squared energy of the final state \( f \), so \( n \)-particle states contribute to the imaginary part of the amplitude only if \( \sqrt{s} \) exceeds the corresponding \( n \)-particle threshold energy. The threshold for producing a state composed of particles with masses \( m_1, m_2, \dots, m_n \) is given by:
\[
s = (m_1 + \cdots + m_n)^2.
\]

To characterize these cuts arising from multi-particle thresholds, let us fix the variable $t$. From the optical theorem in eq. (\ref{eq:unitarity}) we know that $\text{Im}\, T(s, t, u)$ is non-zero along the real axis starting at the first particle threshold up to $\infty$. If the incoming particles are identical this line is $s \in [4m^2, \infty)$.

For simplicity, if we assume that the energy spectrum in the $u$-channel is the same as in $s$, then $\text{Im}\, T(s, t, u)$ will be non-zero for the part of the $u$-axis from $4m^2$ to $\infty$, or in terms of $s$, from $-\infty$ to $-t$. Below any thresholds and for $s>-t$ we have that $\text{Im}\, T(s, t, u) = 0$, then the Schwarz reflection principle tells us that
\[
 T(s^*, t, u) =  T(s, t, u)^*,
\]
for any $s$ and $s^\ast$ within a domain $ \mathcal{D}$ of the complex $s$-plane whose intersection with the real axis lies in a region where the amplitude is purely real.

Using this property, the amplitude can be analytically continued below the branch cut originating from physical thresholds and allows to express its discontinuity across the cut as follows:
\begin{align}
\lim_{\epsilon \to 0^+}  T(s + i\epsilon, t, u) - & T(s - i\epsilon, t, u)
= \lim_{\epsilon \to 0^+} \left[  T(s + i\epsilon, t, u) -  T(s + i\epsilon, t, u)^* \right] \nonumber \\
&= 2i \lim_{\epsilon \to 0^+} \text{Im}\,  T(s + i\epsilon, t, u) \equiv 2i \,\text{Im} \, T(s, t, u),\label{eq:disc}
\end{align}
where the physical amplitude is defined as the value just above the cut.

Similar arguments apply to the physical $t $- and $ u $-channels. These channels exhibit cuts along the positive real $ t $- and $ u $-axes, with branch points located at their respective physical thresholds. Additionally, poles may appear, corresponding to bound states in these channels.

To summarize, in the equal-mass case, the analytic structure of the transition amplitude in the complex $ s $-plane is illustrated in Fig.~\ref{fig:disprel1}. The right-hand cut, extending from $ s = 4m^2 $ to $ +\infty $, arises from the physical threshold in the $ s $-channel. A pole at $ s = s_B $ would indicate a bound state in the $ s $-channel, with mass $ m_B = \sqrt{s_B} $.

The left-hand cut lies along the negative real axis and begins at the branch point $ s = 4m^2 - t - u_0 $, stemming from the physical threshold $ u_0 $ in the $ u $-channel. A bound state in the $ u $-channel would correspond to a  pole at $ s = 4m^2 - t - u_B $, with the bound state at $u=u_B$.

\paragraph{Unequal mass case}
When the kinematics become more involved, additional singularities may appear.  
For instance, in the elastic scattering process
\[
a + b \;\to\; a + b ,
\]
where $m$ and $M$ denote the masses of particles $a$ and $b$ respectively, with $M > m$, the analytic structure of the amplitude is richer than in the equal-mass case.  

In addition to the right- and left-hand cuts already present for equal masses, one finds: a branch cut arising from the $t$-channel, located off the real axis along the circle
\[
|s| \;=\; M^2 - m^2 ,
\]
and a branch cut along the real axis for
\[
s \,\leq\, (M - m)^2 ,
\]
which originates from the $u$-channel.

\subsection{Resonances}\label{subsec:reso}

Coming back to $2\to2$ scattering, when only a single channel is open, the analytic structure consists of just two sheets: the physical (first) sheet and the unphysical (second) sheet. If multiple channels are open, the number of Riemann sheets increases accordingly, as different analytic continuations of the intermediate-state momenta become possible.

By employing equations (\ref{eq:16}) and (\ref{OpticalTh}), one obtains that the scattering amplitude in the second Riemann sheet, $T^{II}(s,t,u)$, equals
\begin{equation}
T^{II}(s,t,u)=\sum_J {P_J(\cos \theta}){t^I_J(s)}/S_J(s)
\end{equation}
which comes from $t^{II}_J(s- i\epsilon) = t^{I}_J (s + i\epsilon)$ followed by $t^{II}_J (s) = t^{I}_J(s)/S_J (s)$,
where $S_J(s)$ is the standard definition of the partial wave projection for the $S$-matrix.

 Hence, the analytic continuation to the second sheet reads:
\begin{equation}
t^{II}_J(s) = \frac{t^I_J(s)}{1 + 2i\,\sigma(s)\,t^I_J(s)},
\end{equation}
where the branch of \( \sigma(s) \) is chosen such that \( \sigma(s^*) = -\sigma(s)^* \) (see equation (\ref{eq:sheetsqrt}) below), ensuring the correct analytic behavior across the cut.

\begin{figure}[ht!]
  \includegraphics[width=0.45\columnwidth]{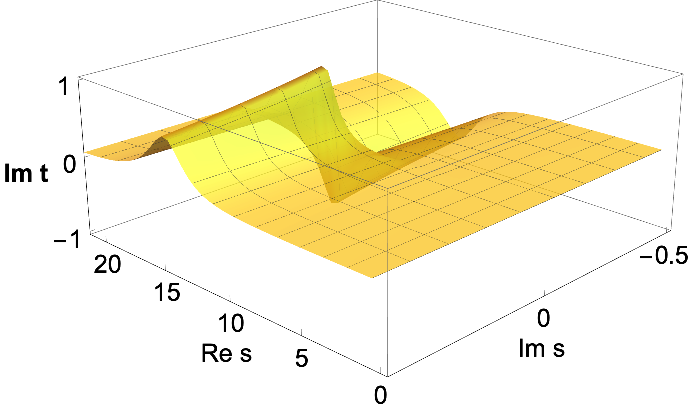}
\includegraphics[width=0.45\columnwidth]{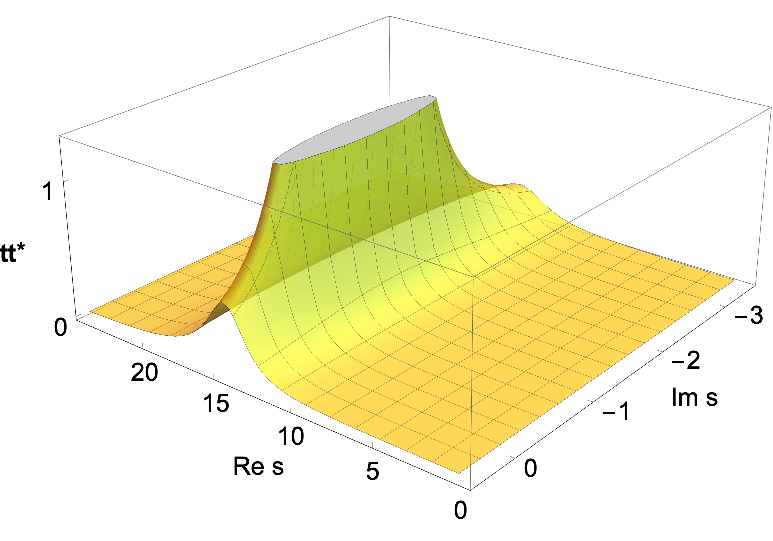}
\caption[Resonances using the IAM extended to the complex $s$ plane]{\label{fig:poleposition2ndsheet}
A resonance using the IAM extended to the complex $s$ plane, with the  parameters, in the notation of \cite{Delgado:2015kxa}, $a=0.95$,
$a_4=-2.5\cdot 10^{-4}$ and $a_5=-1.75\cdot 10^{-4}$. 
The first plot shows the first Riemann sheet, cut along the real axis ${\rm Im}(s)=0$, although it does not leave a pole in this first sheet, it is seen to saturate unitarity (${\rm Im}(s)\simeq 1$ for a certain real $s$ (the scale does not allow to visualize the cut). 
The plot in the bottom line is the extension to the second Riemann sheet and clearly features a pole for negative ${\rm Im} (s)<0$.}
\end{figure}

A resonance will appear at $s=s_R$ as a pole in the second Riemann sheet for the $t_{II}^J(s)$ partial wave whenever 
\begin{equation}
1 + 2i\,\sigma(s_R)\,t_J(s_R)=0
\end{equation}
for the partial wave amplitude in the first Riemann sheet, so that $t_J(s_R)=\frac{i}{2\sigma(s_R)}$.
If one is interested in finding a resonance for real $s$, the real part of the resonance can be found from the saturation of unitarity in eq. (\ref{unitarity}) for the partial waves over the real $s$ axis, \textit{i.e.} $|t_{J}(s_R)| = \frac{1}{\sigma(s_R)}$. These facts can be seen in Fig. \ref{fig:poleposition2ndsheet}. Where the resonance of a unitarized partial wave amplitude, in this case for the HEFT, appears in the second Riemann sheet and the real part of the position of this resonance can be read off from the maximum of the imaginary part over the real axis. 

\section{Causality and Dispersion Relations} \label{sec:DispRel}
Before introducing dispersion relations, and since it is usually overlooked, a comment on the consequences of assuming causal scattering amplitudes is in order. As rigorously established in \cite{Toll:1956cya}, the statement of having a causal scattering amplitude (\textit{i.e.} one that vanishes before any scattering takes place) which respects unitarity, is logically equivalent to requiring that the scattering amplitude be analytic in the complex upper half-plane of the energy variable $E=\sqrt{s}$. This analyticity, in turn, automatically leads to a dispersion relation.  

As a toy example, consider a scattering amplitude in the energy representation $t(E)$, which can be written as the Fourier transform of the scattering amplitude in the time variable $\tau$, denoted $\hat{t}(\tau)$:
\begin{equation}
    t(E) \;=\; \int_{-\infty}^{\infty} \hat{t}(\tau)\, e^{\, i E \tau}\, d\tau .
\end{equation}

If the incoming wave packet hits a pointlike target at $\tau = 0$, causality requires that 
$\hat{t}(\tau) = 0$ for $\tau < 0$. Consequently, since the integrand vanishes at earlier 
times, the lower integration limit can be shifted to $0$. By analytically continuing $t(E)$ 
into the complex plane $E\in \mathbb{C}$, one obtains
\begin{equation}
    t(E) \;=\; \int_{0}^{\infty} \hat{t}(\tau)\, 
    e^{\, i \,\mathrm{Re}(E)\,\tau}\,
    e^{-\, \mathrm{Im}(E)\,\tau}\, d\tau .
    \label{eq:fourier-analytic}
\end{equation}

The exponential damping factor coming from the imaginary part of the energy ensures convergence of the integral in the upper half of the complex $E$--plane. 
As a result, $t(E)$ is analytic for $\mathrm{Im}(E) > 0$ and remains well behaved as $\mathrm{Im}(E) \to +\infty$. 
This analyticity property permits the application of Cauchy’s theorem by closing the contour 
with an infinite semicircle in the upper half-plane, as will be seen next.

\begin{figure}[ht!]
\centering
\includegraphics[width=.7\textwidth]{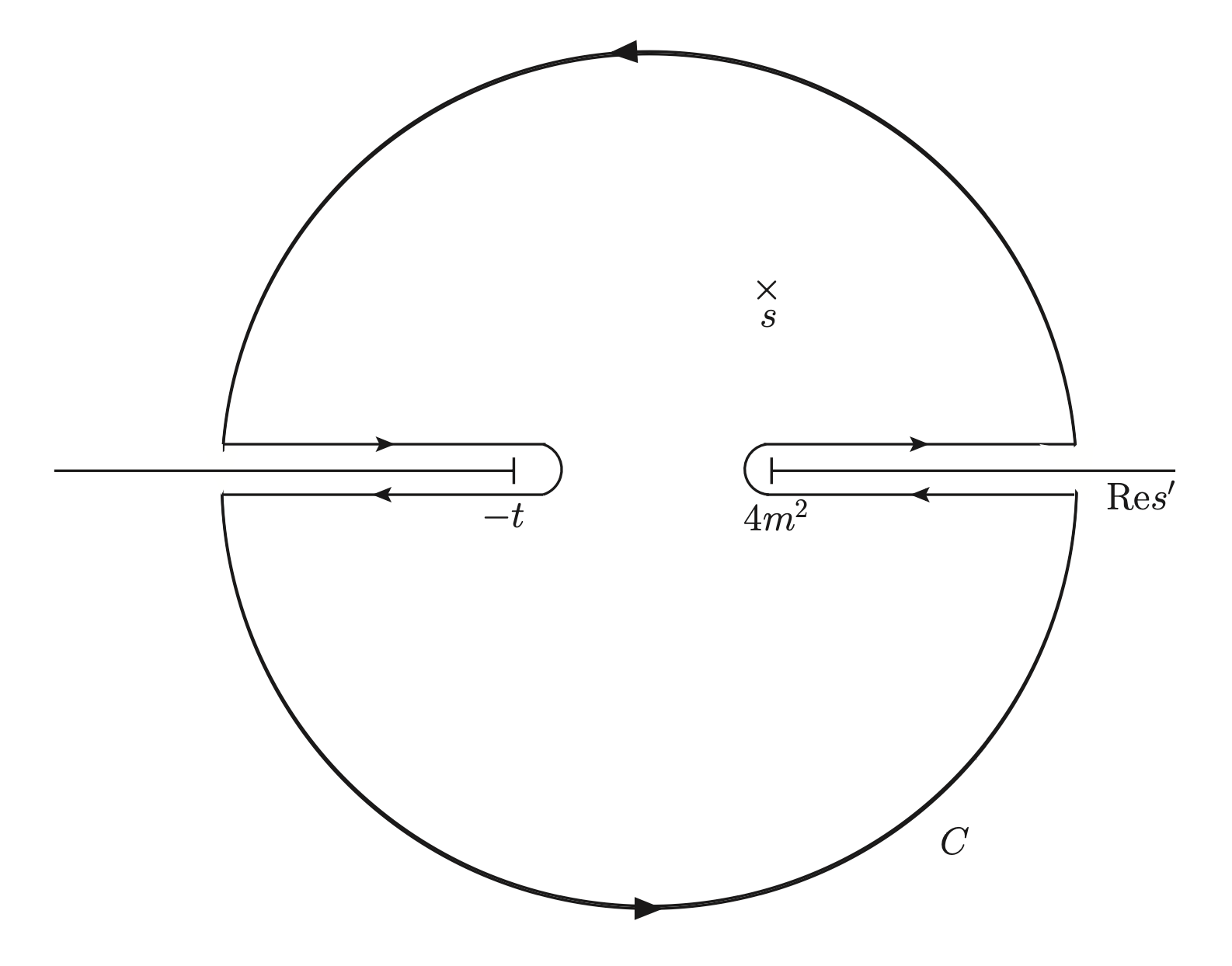}
\caption{Analytic structure of the  $2\to2$ particle scattering amplitude $\M(s,t,u)$ and the contour $C$ for its dispersion relation. Figure from \cite{RuizdeElviraCarrascal:2013tix}.}\label{fig:disprel1}
\end{figure}
For long, dispersion relations have been known to link the imaginary and real parts of ``causal'' functions satisfying Cauchy's theorem in appropriate complex-$E$ (here, $s$) plane regions.  Once we know the analytic structure of a scattering amplitude, we can relate the amplitude at a point with the integral of the amplitude (times appropriate subtraction factors) over a closed contour $C$ around a value $s$. For the case of  $2\to2$ particle scattering with equal masses, we can write down the following simple relation 
\begin{equation}
\M(s,t,u)=\frac{1}{2\pi i}\oint ds'\frac{\M(s',t,u)}{s'-s}\;.
\end{equation}
The contour of integration here can be taken to consist of lines parallel to the left and right cuts of the amplitude, to use an analogous to eq. (\ref{eq:disc}), and closed with a circle of radius $R$ that we will eventually take to infinity (see Fig. \ref{fig:disprel1}). Given that $\M(s,t,u)$ goes to zero faster than $1/s$ as $|s|\to\infty$, then the contribution from this circle vanishes in the limit of $R\to\infty$, leaving us with the dispersion relation 
\begin{equation}
\M(s,t,u)=\frac{1}{\pi}\int_{4m^2}^\infty \frac{\text{Im } \M(s',t,u)}{s'-(s+i\epsilon)}+\frac{1}{\pi}\int_{-\infty}^{-t} \frac{\text{Im } \M(s',t,u)}{s'-(s+i\epsilon)}
\end{equation}
where the $i\epsilon$ prescription is chosen to specify the value of the amplitude above the cut. Remembering the distributional identity
\begin{equation}
\frac{1}{s'-s-i\epsilon}= \text{P.V.}\left( \frac{1}{s'-s} \right)+i\pi\delta(s'-s)\nonumber
\end{equation}
where P.V. denotes the Cauchy's principal value integral, so that we can relate the real and imaginary parts as
\begin{equation}
\text{Re }\M(s,t,u)=\frac{1}{\pi}\text{P.V.}\int_{4m^2}^\infty \frac{\text{Im } \M(s',t,u)}{s'-s}+\frac{1}{\pi}\text{P.V.}\int_{-\infty}^{-t} \frac{\text{Im } \M(s',t,u)}{s'-s}\;.
\end{equation}
We will now move on to the dispersion relations that can be applied to partial waves and come back to complete amplitudes and substracted dispersion relations when dealing with Roy equations in section \ref{sec:Roy}.
\subsection{A dispersion relation for the partial wave amplitude}

\label{InverseAmp}

To exhibit and exploit the analytic structure of the partial wave amplitude $t_{J}(s)$ we will deploy the appropriate dispersion relation (see the book \cite{Oller:2019rej} for a pedagogical account). 
The Cauchy theorem can be applied to any function $f(s)$ which is analytic in a complex-plane domain. 
  \begin{figure}[ht!]
  \centering
\includegraphics[width=0.7\columnwidth]{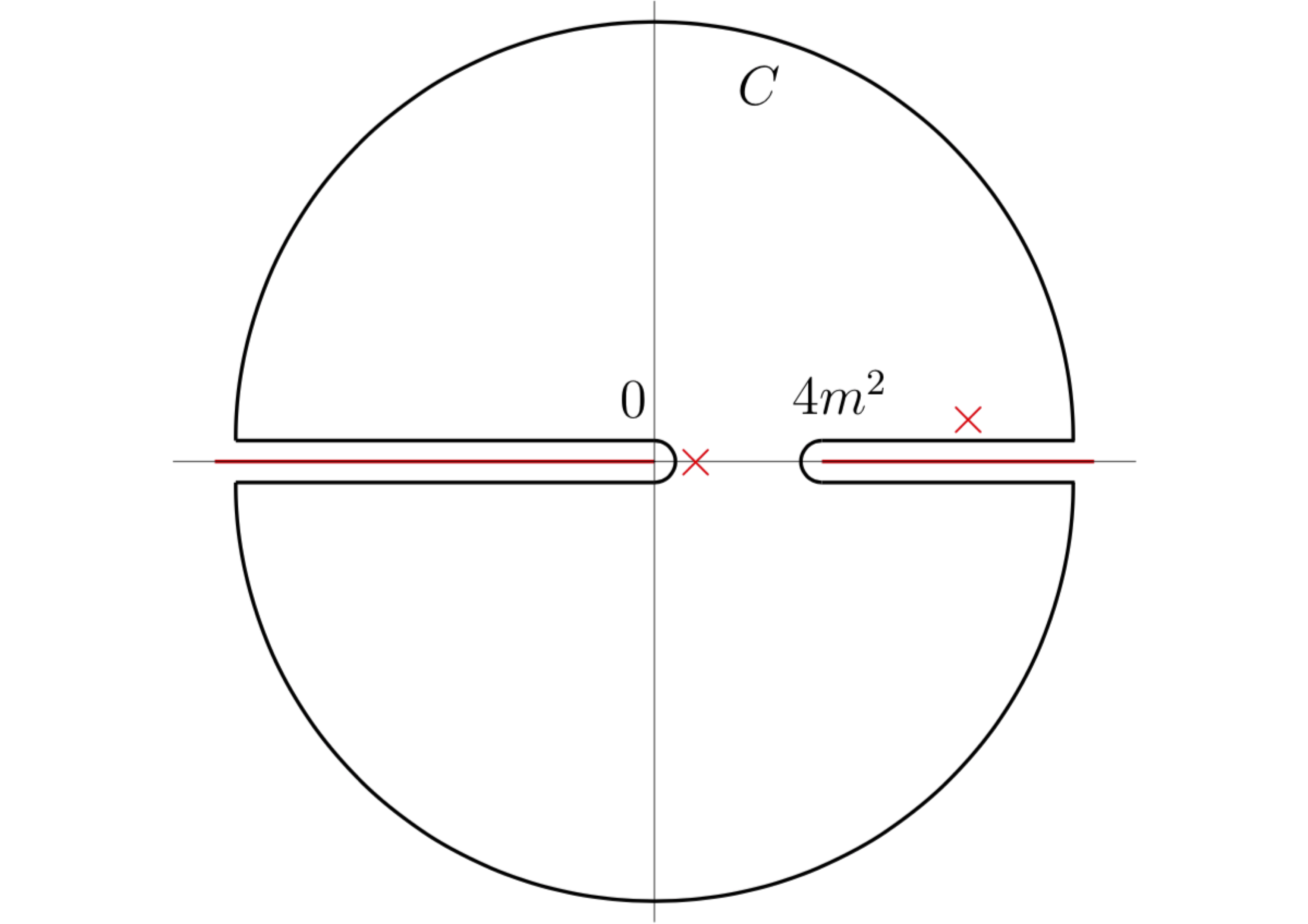}
\caption{Analytic structure of elastic scattering partial waves for scalars of mass $m$ and the contour $C$ in the complex-$s$ plane that will be used to write a dispersion relation. The red lines represent the discontinuity cuts in the partial wave amplitude. The red crosses additionally represent the $n$-th order pole at $z=\epsilon$ and the simple pole at $z=s+i\epsilon$ (with $s>4m^2$) coming from the denominators in eq.~(\ref{I}).}\label{contour}
\end{figure}

Application of the theorem is convenient for the following integral,
\begin{equation}
I(s)\equiv \frac{1}{2\pi i}\int_C dz \frac{f(z)}{(z-\epsilon)^n (z-s)}\label{I}\;,
\end{equation}
where $f(z)$ is taken to have two branch cuts extending from $4m^2$ to $+\infty$ and from $0$ to $-\infty$ (just as the partial wave amplitude $t(s)$), and the contour of integration $C$ is taken as depicted in Fig. \ref{contour}.
In eq. (\ref{I}), $s$ is above the RC (i.e. $s$ stands for $s+i\epsilon$ with $s\geq4m^2$).\par
If the function $f$ is polynomially bounded as $|z|\to \infty$ (which, though not the case for partial wave amplitudes that can diverge exponentially for composite particles~\cite{Llanes-Estrada:2019ktp}, is satisfied for their inverse amplitude in eq.~(\ref{def:inverse}) below, that does fall as $e^{-az}$ with $a$ fixed)
we are able to neglect the contribution to eq. (\ref{I}) coming from the two large semi-circumferences in Fig. \ref{contour}. Due to Schwartz's reflection principle $f(s+i\epsilon)=f^\ast(s-i\epsilon)$ and analogously to eq. (\ref{eq:disc}), we are left in (\ref{I}) with the integrals of the imaginary part of $f$ over the LC and RC,
\begin{eqnarray}
I(s)&=&\frac{1}{\pi}\int_{-\infty}^0 ds'
\frac{\text{Im}f(s')}{(s'-\ep)^n (s'-s)}+\frac{1}{\pi}\int_{4m^2}^\infty dz \frac{\text{Im}f(s')}{(s'-\ep)^n (s'-s)}~,
\end{eqnarray}
where $\text{Im}f(s)$ is the imaginary part of $\lim_{\varepsilon\to 0^+} f(s+i\varepsilon)$ with $s\in$ RC or LC. 
 On the other hand, $I(s)$ equals the sum of its residues coming from the simple pole at $z=s$ and the $n$-th order pole at $z=\epsilon$, with $\ep\in(0,4m^2)$.  In this way we find the $n$-times subtracted dispersion relation for $f(s)$,
\begin{align}
f(s)&=\sum_{k=0}^{n-1}\frac{f^{(k)}(\ep)}{k!}(s-\ep)^k+\nonumber\\
&+\frac{(s-\ep)^n}{\pi}\int_{-\infty}^0 ds' \frac{\text{Im}f(s')}{(s'-\ep)^n (s'-s)}\nonumber+\frac{(s-\ep)^n}{\pi}\int_{4m^2_\pi}^\infty ds' \frac{\text{Im}f(s')}{(s'-\ep)^n (s'-s)}
\end{align}
(which is valid safe at branch points where the multiple derivatives $f^{(k)}$ could fail to exist; this is generally of no concern).

\section{Unitarization Methods for partial waves}\label{sec:pwaunit}

In this section we will briefly summarize, mainly based in the comprehensive reference \cite{Delgado:2015kxa}, the unitarization methods most commonly used for predicting resonances from low energy scattering data.

\subsection{Padé approximants} \label{subsec:Padé}
Historically, Padé approximations to scattering amplitudes have been extensively used both for the strong force and for the electroweak sectors \cite{Bessis:1969uu,Zinn-Justin:1971sac,Filkov:1972jv,Dicus:1990ew,Dobado:1989gv,Dobado:1989gr}. 

In Chiral Perturbation Theory, the partial wave amplitude \( t_J \) admits a Taylor-like expansion in powers of the Mandelstam variable \( s \), modified by logarithmic corrections, for small real \( s \). Suppressing the angular momentum index for brevity, the expansion reads \( t \simeq t_0 + t_1 + \mathcal{O}(s^3) \), where \( t_0 = a + bs \), and each term scales as \( t_i \sim s^{i+1} \) in the low-energy regime.

The $[r,r]$ Padé approximation provides a systematic way to unitarize the perturbative expansion of a partial-wave scattering amplitude. 
Given the expansion up to order $2r$,
\begin{equation}
t(s) = t_0(s) + \cdots + t_{2r-1}(s) \, ,
\end{equation}
the Padé form is
\begin{equation}
t^{[r,r]} = \frac{p_0 + \cdots + p_r}{q_0 + \cdots + q_r},
\end{equation}
with monomials $p_l, q_l$ of $l$-th order in $s$ fixed to reproduce the series up to $2r$,
\begin{equation}
t^{[r,r]} = \frac{p_0 + \cdots + p_r}{q_0 + \cdots + q_r}=t_0(s) + \cdots + t_{2r-1}(s),
\end{equation}
automatically ensuring elastic unitarity in equation (\ref{OpticalTh}) in the massless case (where $t_0\propto s$ and $\sigma(s)=1$).

For example, for the $[1,1]$ massless case one solves
\begin{equation}
t^{[1,1]} = \frac{p_0 + p_1}{q_0 + q_1} = t_0 + t_1 ,
\end{equation}
which gives, matching orders in $s$,
\begin{align}
p_0 = 0, \;\;\;
p_1 = q_0 t_0, \;\;\;
q_1 t_0 + q_0 t_1=0 .
\end{align}
From these relations the explicit Padé result follows:
\begin{equation}
t^{[1,1]} = \frac{t_0^2}{t_0 - t_1} \, .
\end{equation}
This result respects unitarity and has the same form as the IAM amplitude that is derived from a dispersive approach and a NLO chiral truncation that we describe next.

\subsection{The Inverse Amplitude Method} \label{subsec:IAM}

Work from the 1970s highlighted the utility of writing a dispersion relation for the inverse amplitude
for processes such as pion-pion or electroweak Goldstone boson scattering~\cite{Truong:2010wa,Dobado:1989gr,Oller:2020guq,Dobado:1996ps}. 

It has become standard to define the function (likely first introduced by Lehmann~\cite{Lehmann:1972kv})
\begin{equation}
\label{def:inverse}
G(s) \equiv \frac{t_0(s)^2}{t(s)}\ ,
\end{equation}
and write a dispersion relation for it.
This function inherits the analytic structure of \( t_J(s) \), apart from additional poles associated with the zeros of \( t(s) \). At low energies, such zeros—known as Adler zeros~\cite{Adler:1964um}—are particularly relevant in scalar channels.

A third-order subtracted dispersion relation for \( G(s) \) is then constructed, the order being the minimum required to match the polynomial growth dictated by the effective field theory at NLO (\textit{i.e.} $\propto s^2$):
\begin{align}
G(s) &= G(\epsilon) + G'(\epsilon)(s - \epsilon) + \frac{1}{2} G''(\epsilon)(s - \epsilon)^2 + PC(G) \nonumber\\
&\quad + \frac{(s - \epsilon)^3}{\pi} \int_{LC} \frac{\text{Im}\, G(s')}{(s' - \epsilon)^3 (s' - s)}\,ds'
+ \frac{(s - \epsilon)^3}{\pi} \int_{RC} \frac{\text{Im}\, G(s')}{(s' - \epsilon)^3 (s' - s)}\,ds' \,.
\label{disprel}
\end{align}
The first term encountered when analyzing this relation is the Pole Contribution \( PC(G) \), arising from Adler zeros. In the standard Next-to-Leading Order (NLO) IAM, these contributions are neglected since they formally count as Next-to-Next-to-Leading Order (NNLO) on the physical axis~\cite{GomezNicola:2007qj}. However, they can be systematically included, and the induced uncertainty from neglecting them is significantly smaller than the experimental uncertainties in meson-meson scattering data. Moreover, improved versions of IAM that include them differ from the standard result only by \( \mathcal{O}(10^{-3}) \) in the physical region (see below).  
Focusing next on the right-hand cut (RC) due to eq. (\ref{eq:disc}), the IAM treats this contribution exactly, as long as only the elastic two-body channel contributes. Inelastic effects—originating from additional two- or four-body intermediate states—can be addressed in a coupled-channel IAM framework~\cite{Guerrero:1998ei,GomezNicola:2001as}, though with somewhat reduced theoretical control.

As for the subtraction constants \( G(\epsilon) \), \( G'(\epsilon) \), and \( G''(\epsilon) \), three suffice in hadronic applications due to the Froissart-Martin bound~\cite{Froissart:1961ux,Martin:1962rt}, which constrains the asymptotic growth of the cross section. These constants are evaluated using NLO ChPT, a valid approximation near \( s \approx 0 \) where the effective theory is reliable (see the discussion in \cite{Salas-Bernardez:2020hua}). In this low-energy regime, one can safely set \( \epsilon = 0 \) since \( t_0 \) and \( t_1 \) behave essentially as polynomials.

This leads to the intermediate relation:
\begin{align}
\label{intermediateG}
G(s) \equiv \frac{t_0(s)^2}{t(s)} = t_0(s) - t_1(s) + \frac{s^3}{\pi} \int_{LC} \frac{\text{Im}\, G(s') + \text{Im}\, t_1(s')}{s'^3 (s' - s)}\,ds' \,.
\end{align}

Finally, a remarkable simplification emerges when the left-cut integral is approximated by its NLO ChPT expression \( \text{Im}\, G \simeq -\text{Im}\, t_1 \), under which the integral vanishes. The resulting formula,
\begin{equation}
t_{\text{IAM}} \equiv \frac{t_0^2}{t_0 - t_1} \,,
\label{usualIAM}
\end{equation}
defines the standard NLO IAM amplitude. The systematic uncertainties associated with resonance predictions via the IAM, both in pion-pion and electroweak Goldstone boson scattering, have been analyzed in detail in~\cite{Salas-Bernardez:2020hua}.

To properly account for Adler zeros, one may adopt the \textit{Modified Inverse Amplitude Method} (mIAM) introduced in~\cite{GomezNicola:2007qj}. In this approach, the amplitude is expressed as
\begin{equation}
\label{mIAM0}
t_{\rm mIAM} \equiv \frac{t_0^2}{t_0 - t_1 + A_{\rm mIAM}} \,,
\end{equation}
where the correction term \( A_{\rm mIAM} \) incorporates information about the Adler zero position, as computed in chiral perturbation theory, appropriate in the very low-energy regime. Writing \( s_A \simeq s_0 + s_1 + \dots \), its explicit form is
\begin{equation}
\label{mIAM}
A_{\rm mIAM} = t_1(s_0) - \frac{(s_0 - s_A)(s - s_0)}{s - s_A} \left(t_0'(s_0) - t_1'(s_0)\right) \,.
\end{equation}

The appearance of these zeroes below threshold might spoil crossing symmetry and the crossing sum rules \cite{Roskies:1970uj} (Roskies relations) below threshold as seen in \cite{Hannah:1997sm}, modifiying the IAM methods to include these zeroes greatly improves the restoration of crossing symmetry \cite{Hannah:1997sm}. We will comment further on crossing symmetry in subsection \ref{subsec:crossingv}.

Furthermore, the presence of amplitude zeros above threshold—commonly known as Castillejo-Dalitz-Dyson (CDD) zeros—can be accommodated through the modification proposed in~\cite{Salas-Bernardez:2020hua}. The IAM expression is adjusted to
\begin{equation}
\label{modIAM2}
t_{\rm IAM}^{\text{CDD}} =\frac{t_0^2}{t_0 - t_1 + \frac{s}{s - s_C} \text{Re}(t_1)} \,,
\end{equation}
where \( s_C \) denotes the location of the CDD zero, determined by solving the condition
\begin{equation}
t_0(s_C) + \text{Re}\, t_1(s_C) = 0 \,.
\end{equation}
This condition is sufficient to identify the CDD pole in the inverse amplitude for purely elastic scattering.

\subsection{Coupled-channel Inverse Amplitude Method}
\label{subsec:coupledIAM}

The Inverse Amplitude Method  can be naturally extended to the coupled-channel case, provided certain analyticity conditions are met—specifically, that the left-hand and right-hand cuts do not overlap. This condition is typically satisfied when the masses of the particles involved in the various channels are equal or negligible, such as in the high-energy limit of electroweak scalar scattering where all masses can be set to zero~\cite{Delgado:2015kxa, Delgado:2016rtd}.

The formalism for the coupled-channel IAM was originally developed in the context of meson-meson scattering~\cite{Oller:1997ng, Oller:1998hw, Dobado:1996ps, GomezNicola:2001as}, and has since been adapted to other scenarios including electroweak effective theories~\cite{Delgado:2014dxa, Delgado:2015kxa, Pelaez:2015qba}. In this framework, the partial-wave amplitude becomes an $n\times n$ matrix, \(\mathbf{t}\), where \(n\) denotes the number of coupled channels.

The IAM generalization to the matrix case takes the form
\begin{equation}
\mathbf{t}_{\text{IAM}} = \mathbf{t}_0 \left(\mathbf{t}_0 - \mathbf{t}_1\right)^{-1} \mathbf{t}_0 \,,
\end{equation}
which ensures exact unitarity in the elastic regime, provided that the perturbative input satisfies it order by order.

This coupled-channel formulation is particularly valuable when describing processes where multiple scattering channels open up in the same energy region and interact strongly. It has proven effective in capturing resonance behavior and describing data in sectors like the $\pi\pi$–$K\bar{K}$ system~\cite{Oller:1997ng,Oller:1998hw} or in the scalar-isoscalar channel of the electroweak sector~\cite{Delgado:2014dxa,Delgado:2015kxa}. In these treatments one has to be careful that, as said before, right and left cuts do not overlap.

\subsection{The $K$-matrix and the Improved $K$-matrix methods} \label{sec:Km}
Among the earliest and most widely studied unitarization techniques is the $K$-matrix method~\cite{Lippmann:1950zz,Byron:1975nj}. A modern overview of its application to effective field theories, including the electroweak sector, is given in~\cite{Kilian:2014zja}. Limitations of this method and its violation of analyticity conditions can be found in \cite{Truong:1991gv,Qin:2002hk,Masjuan:2008cp}. The $K$-matrix approach ensures unitarity of the $S$-matrix through the relation:
\begin{equation}
S = \frac{1 - i K/2}{1 + i K/2}\;,
\end{equation}
which implies that $S$ is unitary provided the matrix-valued function $K(s)$ is Hermitian. The inverse relation expresses $K$ in terms of $S$ as:
\begin{equation}
K = \frac{i(S - 1)}{1 + \frac{1}{2}(S - 1)}\;.
\end{equation}

While the $S$-matrix is often expressed as a perturbative expansion, here we use the ChPT chiral counting expansion (see for example \cite{Gavela:2016bzc}):
\begin{equation}
S = 1 + S^{(1)} + S^{(2)} + \dots\;,
\end{equation}
such an expansion generally violates unitarity once the expansion is truncated. Alternatively, one may expand $K$ as:
\begin{equation}
K = K^{(1)} + K^{(2)} + \dots\;,
\end{equation}
and substitute it into the non-linear relation for $S$, thereby producing a unitarized expansion that preserves exact unitarity order by order:
\begin{equation}
S = 1 + \tilde{S}^{(1)} + \tilde{S}^{(2)} + \dots\;.
\end{equation}

In the case of elastic scattering, the partial-wave amplitude $t(s)$ is approximated via a real function $a_0(s)$ in the physical region, which clearly does not respect unitarity. Unitarity is then reinstated by defining the $K$-matrix partial wave amplitude:
\begin{equation}
t_K(s) = \frac{a_0(s)}{1 - i \sigma(s) a_0(s)}\;,
\end{equation}
which satisfies the unitarity condition in the physical region:
\begin{equation}
\text{Im}\,t_K = \sigma(s)|t_K|^2 = \frac{a_0^2}{1 + \sigma(s) a_0^2}\;.
\end{equation}

Despite preserving unitarity, this naive $K$-matrix form does not respect the analytic structure in the partial wave amplitude due to the absence of a RC and a LC. This violates microcausality and precludes defining the second Riemann sheet, which is essential for the emergence of resonant poles. Consequently, $K$-matrix amplitudes do not support dynamically generated resonances—a key drawback compared to analytically consistent methods such as the IAM or the N/D approach (treated in section \ref{subsec:ND} below).

This deficiency is well known in hadronic physics, where unitarization methods are routinely employed to describe resonant phenomena such as the $\rho(770)$ or $f_0(980)$~\cite{Oller:1997ng,Oller:1998hw}. The $K$-matrix approach, while robust for low-energy unitarity constraints, is insufficient to account for such resonances without further refinement.

To reintroduce the RC, one defines the analytic function:
\begin{equation}
g(s) = \frac{1}{\pi}\left( C + \sigma(s) \log\left(\frac{-s}{\mu^2}\right) \right)\;,
\end{equation}
where $C$ is a constant and $\mu$ is a reference energy scale. This function is analytic everywhere except for the RC, and its imaginary part in the physical region is:
\begin{equation}
\text{Im}\,g(s) = -\sigma(s)\;.
\end{equation}

Substituting $-i \sigma(s) \rightarrow g(s)$ in the $K$-matrix prescription leads to the \emph{Improved $K$-matrix} (IK) amplitude:
\begin{equation}
t_{\rm IK}(s) = \frac{t_0(s)}{1 + g(s) t_0(s)}\;.
\end{equation}
This amplitude is both unitary and has the correct analytic structure, supporting analytic continuation to the second Riemann sheet and allowing the identification of resonance poles. For instance, setting $a_0(s) = t_{0}(s)$ (the leading-order perturbative amplitude) yields:
\begin{equation}
t_{\rm IK}(s) = \frac{t_{0}(s)}{1 + g(s) t_{0}(s)}\;.
\end{equation}

This form can also be recovered from the twice-subtracted N/D method (see next section) by appropriately choosing subtraction constants \cite{Delgado:2015kxa}. To fully capture the analytic structure of the partial wave amplitude one can incorporate the left-hand cut defining
\begin{equation}
t_{\rm IK}(s) = \frac{t_0(s) +a_L(s)}{1 + g(s)\left[ t_{0}(s) + a_L(s) \right]}\;,
\end{equation}
where $a_L(s)$ is a NLO function incorporating the LC.
The improved $K$-matrix approach is also extendable to coupled-channel scattering. For a multichannel system, one defines the unitarized amplitude matrix in a similar manner:
\begin{equation}
\mathbf{t}_{\rm IK}(s) \equiv \left[1 + \mathbf{g}(s)\,\mathbf{n}(s)\right]^{-1} \mathbf{n}(s)\;,
\end{equation}
where
\begin{equation}
\mathbf{n}(s) \equiv \mathbf{t}_{0}(s) + \mathbf{a}_L(s)\;,
\end{equation}
and $\mathbf{g}(s)$ is the diagonal matrix of loop functions defined analogously to $g(s)$ in the single-channel case and similarly for $\mathbf{a}_L(s)$ with $a_L(s)$.

This formulation restores both unitarity and analyticity, enabling a more physically complete treatment of strongly interacting electroweak boson systems and hadronic resonances alike.

\subsection{The N/D method}\label{subsec:ND}

A third and well known unitarization method is the N/D method \cite{Oller:1998zr}. One starts parameterizing the purely elastic amplitude as (for $J\neq0$)
\be
t_{\text{ND}}(s)\equiv \frac{N(s)}{D(s)}\
\ee
where the numerator function $N(s)$ has only a LC, and the denominator function $D(s)$ has only a RC. This separation guarantees that $t_{\text{ND}}(s)$ exhibits the correct analytic structure, as required by causality.

On the RC, the function $N(s)$ is real, so that $
\text{Im}\, N(s) = 0 $ on the RC,
while on the LC, the denominator remains real: $\text{Im}\, D(s) = 0$.
To satisfy elastic unitarity in the physical region, it is necessary that the imaginary part of the denominator obeys
\[
\text{Im}\, D(s) = - \sigma(s) N(s) \quad \text{(on the RC)},
\]
and correspondingly, the imaginary part of the numerator on the LC satisfies
\[
\text{Im}\, N(s) = D(s)\, \text{Im}\, t(s).
\]
 A convenient normalization can be adopted by setting $D(0) = 1$, which implies
\[
N(0) = t(0),
\]
to ensure the correct value of the amplitude close to the origin. These relations allow one to construct coupled dispersion relations for $N(s)$ and $D(s)$, each integrated over its respective cut:
\begin{eqnarray}\label{N/D}
D(s) & = & 1- \frac{s}{\pi}\int_0^\infty  \frac{N(s')}{s'(s'-s-i\epsilon)}ds' \\ \label{NsobreD2}
N(s) & = & \frac{s}{\pi}\int_{-\infty}^{0}  \frac{D(s') \Imag t(s')}{s'(s'-s-i\epsilon)}ds'    \ .
\end{eqnarray}
More generally, an $n$-times subtracted dispersion relation is required in order to incorporate low-energy dynamics effectively:
\begin{equation}\label{N/D_D_n_subtr}
D(s) = 1 + d_1 s + d_2 s^2 + \dots + d_{n-1} s^{n-1} - \frac{s^n}{\pi} \int_0^\infty \frac{N(s')}{s'^n(s' - s - i\epsilon)}ds'\,.
\end{equation}
The coupled dispersion relations for $N(s)$ and $D(s)$ can, in principle, be solved iteratively. One begins with an approximate function $N_0(s)$ exhibiting the LC—typically a tree-level result—and computes $D_0(s)$ via the dispersion integral over the RC. This yields a first approximation to the amplitude as $t^{(0)}(s) = N_0(s)/D_0(s)$. The process continues by substituting $D_0(s)$ into the second relation to update $N_1(s)$, leading to $t^{(1)}(s) = N_1(s)/D_1(s)$, and so on. With suitable initial conditions, this recursive scheme can converge to the full solution for given subtraction constants. Often, even the simplest approximation $t(s) \simeq N_0(s)/D_0(s)$ provides a reasonable estimate. For instance, with $N_0(s) =  t_{0}(s)$ and using UV and IR cutoffs $\Lambda^2$ and $m^2$, respectively, the result is:
\begin{equation}
D_0(s) = 1 + \frac{t_{0}(s)}{\pi} \log\left( \frac{-s}{\Lambda^2} \right),
\end{equation}
which leads to the approximate amplitude:
\begin{equation}
t(s) \simeq \frac{t_{0}(s)}{1 + \frac{t_{0}(s)}{\pi} \log \left( \frac{-s}{\Lambda^2} \right)}.
\end{equation}
However, this approximation is unsatisfactory compared to, for instance, the IAM presented in section~\ref{subsec:IAM}, since it includes only the RC and omits the LC. Furthermore, it depends on the UV cutoff and does not match the NLO expansion up to $\mathcal{O}(s^2)$, as it neglects loop and chiral corrections \cite{Delgado:2015kxa}.

Incorporating NLO effects into the N/D framework is non-trivial, especially because the one-loop contribution $t_{1}(s)$ contains both LC and RC terms that are individually scale-dependent. The scale-independence of the full amplitude is restored only when combined with the appropriate RGE running of the low-energy constants \cite{Delgado:2015kxa}.

To resolve this, $t_{1}(s)$ is decomposed into two separate, $\mu$-independent functions with definite cut structures, similarly to the $K$-matrix method by defining:
\begin{equation}\label{gB}
g(s) = \frac{1}{\pi} \left( \frac{B(\mu)}{D+E} + \sigma(s) \log\left( \frac{-s}{\mu^2} \right) \right).
\end{equation}
Where $B(\mu)$ is a running constant and $D$ and $E$ are constants that depend on isospin and angular momentum numbers.
This function is analytic except along the RC and satisfies $\Imag g(s) = -\sigma(s)$ on that cut. Then we define:
\begin{align}
a_L(s) &= \pi g(-s) D s^2, \\
a_R(s) &= \pi g(s) E s^2,
\end{align}
so that $t_1(s)=a_R(s)+a_L(s)$. This decomposition may not be possible for certain channels \cite{Delgado:2015kxa}.
In this way, the full amplitude at NLO becomes:
\begin{equation}
t(s) =t_{0}(s) + a_L(s) - [t_{0}(s)]^2 g(s) + \mathcal{O}(s^3).
\end{equation}

A first guess for the numerator approximation is:
\begin{equation}\label{numeratordef}
N_0(s) = t_{0}(s) + a_L(s),
\end{equation}
which contains the LC, encodes chiral dynamics, and is $\mu$-independent.

The drawback is that the dispersion integral for $D_0(s)$ now requires better UV behavior due to the $s^2$ term. Three subtractions suffice to regularize it:
\begin{equation}
D_0(s) = 1 + d_1 s + d_2 s^2 - \frac{s^3}{\pi} \int_0^\infty \frac{ds'\, [t_0(s') + a_L(s')]}{s'^3(s' - s - i\epsilon)}.
\end{equation}
Then, the partial wave becomes:
\begin{equation}
t_{\rm ND}(s) = \frac{N_0(s)}{D_0(s)}.
\end{equation}
With the denominator being:
\begin{equation}
D_0(s) = 1 - \frac{a_R(s)}{t_0(s)} + \frac{\pi}{2} [g(s)]^2 D s^2.
\end{equation}

This formulation has several advantages. It is UV- and IR-finite, $\mu$-independent, satisfies exact elastic unitarity:
\begin{equation}
\Imag t_{\rm ND}(s) = \sigma (s) \lvert t_{\rm ND}(s) \rvert^2,
\end{equation}
and reproduces the NLO amplitude up to $\mathcal{O}(s^2)$:
\begin{equation}
t_{\rm ND}(s) = t_0(s) + t_1(s) + \mathcal{O}(s^3).
\end{equation}
These properties render the N/D method comparably robust to the IAM. Furthermore, it can be explicitly demonstrated that the N/D amplitude reduces to the IAM result in the limit where the left-cut contribution is subdominant, i.e., when $a_L(s) \ll t_0(s)$.


\subsection{Resonances for the different methods}
As previously discussed, the IAM, N/D, and IK unitarization methods can yield poles on the second Riemann sheet below the real axis they can be interpreted as dynamical resonances when sufficiently close to the real axis in the complex $s$-plane (narrow resonances).

In the amplitudes built from ChPT at finite order, the non-trivial analytic structure originates from logarithmic terms produced by loop calculations, defined on the first Riemann sheet as $\log(z) = \log|z| + i \arg(z)$, with $\arg(z)$ cut along the negative real axis. To extend to the second Riemann sheet (one of the infinite sheets of the logarithm's Riemann surface), one uses:
\[
\log^{II}(-z) = \log|z| + i[\arg(z) - \pi],
\]
and searches for zeros of the second-sheet amplitude denominators, $t^{II}(s)$ or $\mathbf{t}^{II}(s)$ in coupled channels, as in~\cite{Delgado:2013loa}. This definition coincides with the usual convention that defines the only two sheets on the Riemann surface that define uniquely the square root appearing in the kinematical factor. To see this notice that
\[
\sqrt{z}^{I}=e^{\frac{1}{2}(\log|z| + i[\arg(z)])}
\]
and
\begin{equation}
\sqrt{z}^{II}=e^{\frac{1}{2}(\log|z| + i[\arg(-z)-\pi])}=e^{\frac{1}{2}(\log|z| + i[\arg(z)-2\pi])}=-\sqrt{z}^{I}\label{eq:sheetsqrt}
\end{equation}
for $s$ in the upper-half complex plane. 

As in section \ref{subsec:reso}, for elastic amplitudes, the second-sheet continuation can be written:
\[
t^{II}(s) = \frac{t(s)}{1 + 2i\sigma(s) t(s)},
\]
so that resonances satisfy:
\[
t(s_R) - \frac{i}{2\sigma(s_R)} = 0.
\]

Given the pole position $s_R = M^2 - i \Gamma M$, the width-to-mass ratio is $\gamma = \Gamma / M$, and $s_R = |s_R| e^{-i\theta}$ with $\theta = \tan^{-1} \gamma$.

The resonance condition takes a specific form for each method. For the IAM:
\[
t_0(s_R) - t_1(s_R) + 2i \sigma(s_R) [t_0(s_R)]^2 = 0,
\]
for the N/D method:
\[
t_0(s_R) - a_R(s_R) + \frac{1}{2} g(s_R) t_0(s_R) a_L(-s_R) + 2i \sigma(s_R) t_0(s_R)[t_0(s_R) + a_L(s_R)] = 0,
\]
and for the IK method:
\[
t_0(s_R) - a_R(s_R) + g(s_R) t_0(s_R) a_L(s_R) +2i \sigma(s_R) t_0(s_R)[t_0(s_R) + a_L(s_R)] = 0.
\]

These equations are renormalization-scale independent, as all $\mu$-dependence cancels. Although the methods differ in form, their predictions converge when the LC contributions is small $a_L(s_R) \ll 1$, since $t_1(s_R) = a_R(s_R) + a_L(s_R)$.
\subsection{On crossing-symmetry violation of the partial wave expansion}\label{subsec:crossingv}

Although the underlying effective field theory is crossing symmetric by virtue of Lorentz invariance and locality, the unitarization procedures employed---such as the IAM---treat the analytic structures associated with the left- and right-hand cuts asymmetrically. This raises questions regarding the extent to which crossing symmetry is preserved in the resulting amplitudes~\cite{Cavalcante:2001yw,Cavalcante:2002bit}. In \cite{Guo:2007ff} it is shown, in the large $N_c$ limit, that partial wave amplitudes do respect crossing symmetry and that standard unitarization techniques of these amplitudes spoil it (see also \cite{Guo:2006kj}). 

Crossing symmetry becomes manifest in the decomposition of $SU(2)$ Goldstone boson scattering into isospin amplitudes, $T_I$:
\begin{align} \label{isospindec}
T_0(s,t) &= 3T(s,t,u) + T(t,s,u) + T(u,t,s), \nonumber\\
T_1(s,t) &= T(t,s,u) - T(u,t,s), \nonumber\\
T_2(s,t) &= T(t,s,u) + T(u,t,s),
\end{align}
where a single analytic function $T(s,t,u)$ is evaluated in different kinematic regions by permuting the Mandelstam variables as in section \ref{sec:crossing}.

If $T(s,t,u)$, $T(t,s,u)$, and $T(u,t,s)$ were unrelated complex functions, the isospin amplitudes could be reconstructed independently by applying unitarization (e.g., IAM) to each isospin channel and then inverting the system in eq.~\eqref{isospindec}. However, since these are analytic continuations of a single function, crossing symmetry imposes nontrivial constraints that are not automatically enforced by unitarization.

The Roy equations~\cite{Roy:1971tc}, that we will describe next in section \ref{sec:Roy}, provide a framework to encode these constraints through dispersion relations involving partial waves of different angular momentum. Yet, unitarization methods at NLO, such as the IAM, are typically truncated at $J < 2$, limiting their ability to test or reconstruct the full crossing-symmetric amplitude. Thus, although IAM reliably captures the low-lying partial waves and resonance behavior, it lacks the necessary coverage to make conclusive statements about crossing symmetry.

Attempts to test crossing symmetry directly in the physical region using eq.~\eqref{isospindec} have revealed significant violations~\cite{Cavalcante:2001yw}, but such discrepancies are not unique to the IAM and persist in any reasonable parametrization of the experimental data. This suggests that either higher-spin contributions are essential to test crossing symmetry accurately, or that the data itself may exhibit inconsistencies with crossing symmetry.

A more controlled test involves the Roskies relations~\cite{Roskies:1970uj} or the more general Balachandran-Nuyts relations \cite{Balachandran:1968zza,Balachandran:1968rj}, which are integral constraints valid below threshold ($s < 4m^2$) where only a few partial waves contribute. The main idea used to obtain these relations is to decompose all isospin amplitudes $T_I$ in terms of three totally symmetric functions in $s$, $t$ and $u$ and expand them in terms of a complete basis of orthonormal functions for all functions in the Mandelstam triangle ($s,t,u>0$ below any thresholds) depicted in Figure \ref{regionm}. Using the crossing relations among isospin amplitudes and the orthonormality of the basis functions one arrives at relations for the $t_{IJ}(s)$ partial waves such as
\begin{equation}
   \int_0^{4 m^2} (4m^2-s)(2 t_{00} (s)-5t_{20}(s))ds=0,
\end{equation}
which test the how much the unitarization technique used respects crossing symmetry.

These relations were evaluated using IAM amplitudes (with approximate treatment of Adler zeros in the absence of the mIAM at the time), and found to be satisfied to within $\mathcal{O}(1\%)$ for the lowest-order relations~\cite{Nieves:2001de}. Since this kinematic region is well-described by chiral perturbation theory, the small deviations are more indicative of EFT uncertainties than failures of the unitarization scheme.

\section{Roy Equations}\label{sec:Roy}

We now turn to the twice-subtracted Roy equations~\cite{Roy:1971tc}, a framework that stands out for its rigorous incorporation of both analyticity and crossing symmetry. These equations constitute an infinite system of coupled integral relations for the partial-wave amplitudes \( t_{IJ}(s) \), organized by isospin \( I \) and angular momentum \( J \). They can be understood as dispersion relations evaluated away from the forward limit, supplemented by the imposition of crossing symmetry between the \( s \)- and \( t \)-channels.

A key advantage of the Roy formalism lies in its treatment of the LC. While in most unitarization approaches discussed earlier—such as the N/D method (subsection \ref{subsec:ND}), which neglects the LC, and the IAM (subsection \ref{subsec:IAM}), which approximates it via Chiral Perturbation Theory —the LC is either disregarded or estimated perturbatively, the Roy equations recast it as a series of integrals over the physical region. This allows for a data-driven and systematically controlled reconstruction of the LC, making the method particularly powerful in the low-energy regime.

Thanks to this structure, the Roy equations can be analytically continued into the complex \( s \)-plane, enabling a robust identification of resonance poles on the second Riemann sheet. They have thus become a cornerstone in the precise analysis of low-energy hadronic scattering and are yet to be used in the EW scalar sector.

Originally derived in~\cite{Roy:1971tc}, the Roy equations were extensively studied throughout the 1970s~\cite{Pennington:1973xv,Basdevant:1973ru,Froggatt:1977hu,Ananthanarayan:2000ht,Colangelo:2001df,Buettiker:2003pp,Kaminski:2006qe}. Also, the Roy equations can be extended to unequal mass cases and are referred to as the Roy-Steiner equations \cite{Hite:1973pm} (see \cite{Hoferichter:2015hva} for a comprehensive review). In more recent decades, their relevance has been renewed by advances in ChPT and the availability of high-precision low-energy $\pi\pi$ scattering data~\cite{Colangelo:2000jc,Garcia-Martin:2011iqs}. They have proven instrumental in refining experimental analyses, eliminating ambiguous solutions~\cite{Descotes-Genon:2006sdr}, testing predictions from ChPT~\cite{Gasser:1983yg,Colangelo:2001df}, and determining the $\pi\pi$ scattering amplitude with unprecedented accuracy~\cite{Caprini:2005zr,Garcia-Martin:2011nna,Pelaez:2015qba}. 

Basically, Roy equations use a twice substracted dispersion relation in the $s$ variable for the scattering amplitude $T(s,t,u)$ transforming the LC integral into $u$-channel RC integrals. This $u$-channel RC integral is finally transformed into a RC integral via crossing relations. By fixing the subtraction constants and projecting into partial waves one finds Roy equations (see \cite{RuizdeElviraCarrascal:2013tix}). Roy equations are only valid for values of $t$ inside the above-mentioned Lehmann ellipse at fixed $s$ \cite{Lehmann:1972kv}. The generic structure of the Roy equations is:
\begin{equation}
\text{Re} \, t_{IJ}(s) = ST_{IJ}(s) + \sum_{J'=0}^{\infty} \sum_{I'=0}^{2} \int_{4m_\pi^2}^{\infty} ds' \, K_{II'JJ'}(s', s) \, \text{Im} \, t_{I'J'}(s'),
\end{equation}
where the kernels \( K_{II'JJ'}(s', s) \) are known, and the subtraction terms \( STI^J(s) \) depend on a few low-energy constants, typically the $S$-wave scattering lengths. The infinite sums are truncated in practical implementations—commonly at \( J < 2 \)—and the contributions from higher partial waves and high energies are bundled into ``driving terms,'' treated as external inputs.
 For example \cite{RuizdeElviraCarrascal:2013tix}, the equation for the $I=0$ $S$-wave reads:
\begin{align}
\text{Re} \, t_{00}(s) &= a_0^0 + \frac{(2a_0^0 - 5a_2^0)(s - 4m_\pi^2)}{12m_\pi^2} \notag \\
&\quad + \sum_{J'=0,1} \sum_{I'=0}^2 \int_{s_0}^{\infty} ds' \, K_{0I'0J'}(s', s) \, \text{Im} \, t_{I'J'}(s') + DT_0^0(s),
\end{align}
with analogous expressions for \( t_{20}(s) \) and \( t_{11}(s) \). These involve subtraction terms dictated by the $S$-wave scattering lengths \( a_0^0 \) and \( a_2^0 \), as well as driving terms \( DT_I^J(s) \) accounting for neglected contributions beyond the truncation cutoff \( s_0 \). The subtractions render the equations especially suitable for probing low-energy physics, where analytic control is most complete and experimental data are most precise.

Roy equations provide a unique, systematically improvable tool to connect experimental data, EFT predictions, and analytic constraints in a unified framework.

Recent developments have broadened the scope of Roy-like equations and their applications \cite{Pelaez:2024uav}. Another example of a variation of Roskies and Roy equations is found in \cite{Cao:2023ntr}, where a $\sigma$ bound state is analyzed for unphysical pion masses. This analyses are crucial for testing lattice data (where pions are typically heavier than the physical one).

\subsection{Left cut for partial waves}
A new representation of partial waves over the LC that could be tested against Roy equations was obtained in \cite{Salas-Bernardez:2020hua}. This representation tries to connect the LC of a partial wave with the amplitude and its derivatives on the physical $t$- and $u$-channels. We will briefly describe this representation:
The partial wave projection along the left-hand cut ($s \leq 0$) is given by eq. (\ref{eq:pwa}):
\begin{equation}
t_{IJ}(s) = \frac{1}{32\pi\eta} \int_{-1}^{+1} dx \, P_J(x) \, T_I(s,t(x),u(x)) \label{lcpartialwv}
\end{equation}
with the Mandelstam variables parametrized as $t(x) = (2m^2 - s/2)(1 - x)$ and $u(x) = (2m^2 - s/2)(1 + x)$.

\begin{figure}[ht!]
\centering
\includegraphics[width=0.6\columnwidth]{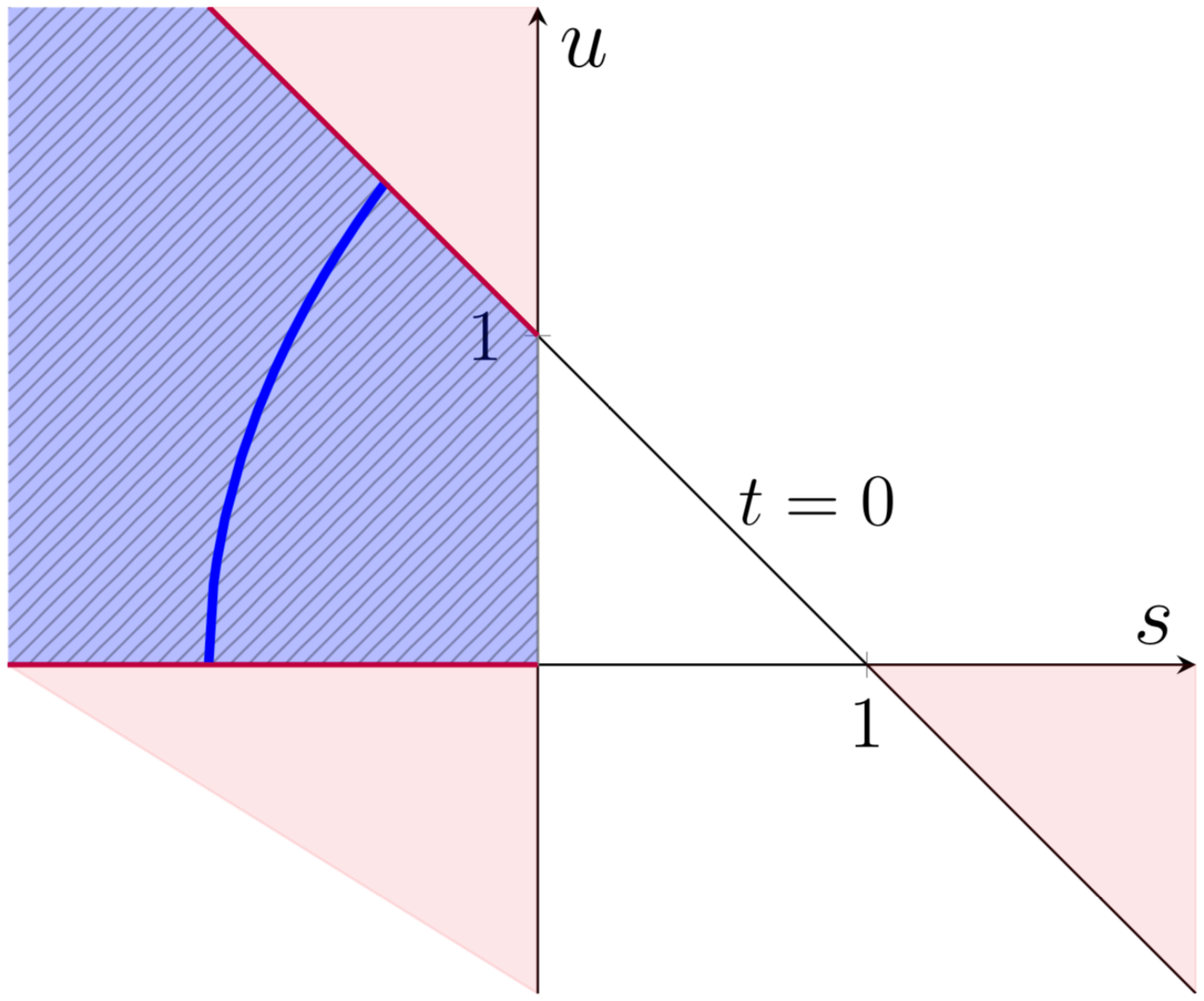}
\caption[Mandelstam plane for two identical particles]{
Mandelstam plane for $2 \to 2$ scattering of identical particles (units: $4m^2$, with $t = 1 - s - u$). The three physical regions (misty rose) extend outward from the triangle vertices. The unphysical integration domain for the left cut (patterned, purplish) is traversed by the arc (thick blue line) corresponding to $x \in [-1, 1]$. Endpoints lie on the $t$- and $u$-channel physical cuts (velvet).
}
\label{region}
\end{figure}

Following Mandelstam's prescription, we assume the existence of a single analytic function $T(s,t,u)$ that governs the amplitude in all three physical regions as in section~\ref{sec:crossing}. These three physical regions are shaded (misty rose) in Fig.~\ref{region}. The integration domain in eq.~\eqref{lcpartialwv} corresponds to a line that traverses the unphysical region of the Mandelstam plane (patterned, purplish), ending on the physical $t$- and $u$-channel thresholds (velvet-colored).

At the endpoints of the integral, the amplitude evaluates to values on the physical right-hand cuts of the $t$- and $u$-channels:
\[
\begin{cases}
T(s,x = +1) = T(s, 0, 4m^2 - s) = T_u(4m^2 - s, 0, s) \,, \\
T(s,x = -1) = T(s, 4m^2 - s, 0) = T_t(4m^2 - s, s, 0) \,.
\end{cases}
\]
Since $s < 0$ and we evaluate $T(s+i\epsilon,x)$, the corresponding $t$ and $u$ arguments lie just below their respective right-hand cuts.

The partial wave can be expressed in terms of auxiliary moments (omitting isospin indices):
\begin{equation}
\psi_J(s) \equiv \int_{-1}^{+1} dx \, x^J \, T(s,t(x),u(x)) \,. \label{si}
\end{equation}
Using Cauchy's inequalities for analytic functions, we integrate by parts an infinite number of times:
\begin{align}
\psi_J(s) &= \sum_{n=0}^\infty \frac{(-1)^n J!}{(n+1+J)!} \Bigg[
\left. \frac{\partial^n T(s,x)}{\partial x^n} \right|_{x = +1} 
+ (-1)^{n+J} \left. \frac{\partial^n T(s,x)}{\partial x^n} \right|_{x = -1} 
\Bigg] \,.
\label{series}
\end{align}

The endpoint derivatives are related to those of the amplitudes in the $u$- and $t$-channels:
\begin{align}
\left. \frac{\partial^n T(s,x)}{\partial x^n} \right|_{x=+1} 
&= (2m^2 - s/2)^n \left. \frac{\partial^n T_u(u,0,s)}{\partial u^n} \right|_{u = 4m^2 - s} \,, \\
\left. \frac{\partial^n T(s,x)}{\partial x^n} \right|_{x=-1} 
&= (s/2 - 2m^2)^n \left. \frac{\partial^n T_t(t,s,0)}{\partial t^n} \right|_{t = 4m^2 - s} \,.
\end{align}

These relations hold provided $s \neq 0$, to avoid the branch points from the $u$- and $t$-channels overlapping at $x = \pm1$. Since $T$ is analytic on the first Riemann sheet, and its derivatives are bounded, the expansion converges.

To convert from monomials to Legendre polynomials, we write:
\begin{equation}
P_J(x) = 2^J \sum_{k=0}^J \binom{J}{k} \binom{\frac{J + k - 1}{2}}{J}_g x^k \,,
\end{equation}
where the generalized binomial coefficient is
\begin{equation}
\binom{\alpha}{k}_g = \frac{\alpha(\alpha - 1)\cdots(\alpha - k + 1)}{k!} \,.
\end{equation}

Substituting into eq.~\eqref{lcpartialwv}, the partial wave becomes
\begin{align}
t_J(s) &= \frac{2^J}{32\pi\eta} \sum_{k=0}^J \binom{J}{k} \binom{\frac{J+k-1}{2}}{J}_g
\sum_{n=0}^\infty \frac{(-1)^n k!}{(n+1+k)!} \times \nonumber \\
&\quad \times \left(
\left. \frac{\partial^n T(s,x)}{\partial x^n} \right|_{x=+1} 
+ (-1)^{n+J} \left. \frac{\partial^n T(s,x)}{\partial x^n} \right|_{x=-1}
\right) \,.
\label{leftcutt}
\end{align}

This expansion expresses the partial wave over an unphysical region (the left cut) in terms of physically accessible quantities at $u$- and $t$-channel thresholds. Studies such as~\cite{Oehme:1956zz} provide dispersion relations for forward amplitude derivatives and could be instrumental in future applications.


\section{Conclusions}

In this work, we have presented a comprehensive review of unitarization methods that extend the applicability of Effective Field Theories into resonant regions. These techniques (the Inverse Amplitude Method (IAM), the improved K-matrix formalism, and the N/D approach) fix the violation of perturbative unitarity that typically arises in hadronic and electroweak scattering processes at energies beyond the threshold of naive EFT expansions.

This review has emphasized how these methods, rooted in analyticity, unitarity, and causality, enable the resummation of low-energy expansions while dynamically generating resonance phenomena. Moreover, we have highlighted the unique role of dispersive techniques, particularly Roy equations, in providing a rigorous framework that enforces crossing symmetry and enables precise predictions from experimental data and chiral input.

While IAM and related methods are successful in reproducing known resonances and offer practical unitarization schemes, their approximate treatment of the left-hand cut and potential crossing symmetry violations remain open for improvement. The Roy framework stands out by offering a systematic and model-independent tool for overcoming these limitations—yet its application to the electroweak sector remains largely unexplored.

These findings suggest a promising path forward: combining non-perturbative unitarization with dispersive approaches to achieve greater theoretical control over EFT predictions. This will be especially relevant for interpreting future high-precision data and probing indirect signatures of new physics beyond the Standard Model.

\bibliographystyle{JHEP}
\bibliography{references.bib}
\end{document}